\documentclass[a4paper,preprintnumbers,floatfix,superscriptaddress,pra,twocolumn,longbibliography]{revtex4-1}

\usepackage{amsmath, amsthm, amssymb}
\usepackage{graphicx}
\usepackage{dcolumn}
\usepackage{bm}
\usepackage{bbm}
\usepackage{color}
\usepackage[usenames,dvipsnames]{xcolor}
\usepackage{hyperref}
\usepackage{enumerate}
\usepackage{ulem}
\usepackage{textcomp}

\newcommand{\figref}[1]{Fig.~\ref{#1}}

\newcommand{\appref}[1]{App.~\ref{#1}}

\newcommand{\ket}[1]{| #1 \rangle}
\newcommand{\bra}[1]{\langle #1 |}
\newcommand{\braket}[2]{\langle #1|#2\rangle}

\newcommand{\Tr}{\operatorname{Tr}}

\newcommand{\ba}{\begin{eqnarray}}
\newcommand{\ea}{\end{eqnarray}}

\begin{document}

\title{MHz-rate semi-device-independent quantum random number generators \\ based on unambiguous state discrimination}


\author{Jonatan Bohr Brask}\thanks{These authors contributed equally to this work.}\affiliation{Group of Applied Physics, Universit\'e de Gen\`eve, 1211 Gen\`eve, Switzerland}
\author{Anthony Martin}\thanks{These authors contributed equally to this work.}\affiliation{Group of Applied Physics, Universit\'e de Gen\`eve, 1211 Gen\`eve, Switzerland}
\author{William Esposito}\affiliation{Group of Applied Physics, Universit\'e de Gen\`eve, 1211 Gen\`eve, Switzerland}
\author{Raphael Houlmann}\affiliation{Group of Applied Physics, Universit\'e de Gen\`eve, 1211 Gen\`eve, Switzerland}
\author{Joseph Bowles}\affiliation{Group of Applied Physics, Universit\'e de Gen\`eve, 1211 Gen\`eve, Switzerland}
\affiliation{ICFO-Institut de Ciencies Fotoniques, Mediterranean Technology Park, 08860 Castelldefels (Barcelona), Spain}
\author{Hugo Zbinden}\affiliation{Group of Applied Physics, Universit\'e de Gen\`eve, 1211 Gen\`eve, Switzerland}
\author{Nicolas Brunner}\affiliation{Group of Applied Physics, Universit\'e de Gen\`eve, 1211 Gen\`eve, Switzerland}

\begin{abstract}
An approach to quantum random number generation based on unambiguous quantum state discrimination (USD) is developed. We consider a prepare-and-measure protocol, where two non-orthogonal quantum states can be prepared, and a measurement device aims at unambiguously discriminating between them. Because the states are non-orthogonal, this necessarily leads to a minimal rate of inconclusive events whose occurrence must be genuinely random and which provide the randomness source that we exploit. Our protocol is semi-device-independent in the sense that the output entropy can be lower bounded based on experimental data and few general assumptions about the setup alone. It is also practically relevant, which we demonstrate by realising a simple optical implementation achieving rates of 16.5\,Mbits/s. Combining ease of implementation, high rate, and real-time entropy estimation, our protocol represents a promising approach intermediate between fully device-independent protocols and commercial QRNGs.
\end{abstract}

\maketitle


Many tasks in modern science and technology make use of random numbers, including Monte Carlo simulation, statistical sampling, cryptography, and gaming applications~\cite{Hayes2001}. In general, a good random number generator is desired to produce output with a high entropy and at a high rate. For applications requiring security, such as cryptography and gambling, the randomness must be certified relative to any untrusted parties. Due to the inherent randomness in quantum physics, in recent years, intense effort has been devoted to extracting randomness from quantum systems, and quantum random number generation (QRNG) devices are now commercially available~\cite{Herrero-collantes2014,Bera2016}. 

QRNG can be implemented in a simple setup, exploiting the randomness in a quantum measurement. For example, one may send a single photon onto a balanced beam splitter and detect the output path~\cite{rarity1994,stefanov2000,jennewein2000}. Other variants measure the arrival time of single photons~\cite{dynes2008,wahl2011,nie2014,stipcevic2007,Stipcevic2015}, the phase noise of a laser~\cite{qi2010,uchida2008,abellan2014}, vacuum fluctuations~\cite{gabriel2010,symul2011}, and shot-noise in mobile phone cameras~\cite{sanguinetti2014}. However, the principle is essentially the same. The device produces a string of raw bits, which in general contains some amount of randomness but is not perfectly random. In order to extract a final (almost) perfectly random bit string, one uses a randomness extractor~\cite{nisan1999}. The correct use of such extractors requires a good estimate of the entropy of the raw data. This can be obtained via detailed theoretical modelling of the setup~\cite{frauchiger2013,ma2013}, but this is usually cumbersome and challenging. Moreover, any mismatch between the model and the implementation, or the instability of the device may jeopardize the security of the protocol. 

It turns out that these problems can be circumvented via the so-called device-independent (DI) approach to randomness certification. In a setup violating a Bell inequality, the entropy of the output data can be certified without any detailed knowledge of the physical implementation~\cite{colbeckPhD,pironio2010}; see \cite{Acin2016} for a review. This provides a highly reliable and secure form of randomness, as it allows the physical devices to be completely untrusted and is thus robust against imperfection in implementation. However, it is technologically extremely challenging to realise as it requires Bell-inequality violation with no post-selection. So far, only proof-of-principle experiments were reported~\cite{pironio2010,christensen2013}, achieving very low bit rates.

More recently, an intermediate approach termed semi-DI has been discussed, exploring the trade-off between ease of implementation and strong security \cite{pawlowski2011,li2011,li2012,bowles2013,Woodhead2015}. Usually based on a prepare-and-measure setup (hence avoiding the complication of a Bell test), these schemes gain ease of implementation by introducing some level of trust in the devices used. Still, they require only general assumptions about the physical implementation, such as bounded dimension~\cite{lunghi2015,canas2014,Mironowicz2016}, trusted measurement devices \cite{vallone2014,Marangon2017,Cao2016,Xu2016}, or a trusted source~\cite{Cao2015}. While significant progress has been achieved, it is fair to say that the right balance between simplicity, performance, and security has yet to be identified. 

Here, we explore a novel approach to quantum random number generation, based on unambiguous quantum state discrimination (USD). Specifically, a quantum system is prepared in one out of two quantum states which are non-orthogonal and hence cannot be distinguished with certainty. However, by performing a USD measurement, the two states can be unambiguously distinguished (i.e.~without false positives), at the price of having a certain minimal rate of inconclusive events \cite{Ivanovic1987,Dieks1988,Peres1988}; see also \cite{Chefles2000,Barnett2009}. The occurrence of these inconclusive events must be genuinely random (if not, the states could be distinguished better), and this is the source of quantum randomness that we use. Our protocol is semi-DI in the sense that the output entropy can be lower bounded based on experimental data and a few general assumptions about the setup. The concept is general, and can thus be implemented in a variety of physical systems. We have implemented the protocol in a simple optical setup using time-bin or photon number encoding. Our setup features only standard components and achieves a rate of 16.5\,Mbits/s, comparable with commercial QRNGs. Hence our protocol combines high performance and ease of implementation with the possibility for the user to verify the generation of certified quantum randomness in real-time. 

\begin{figure}[t!]
  \centering
  \includegraphics[width=0.99\columnwidth]{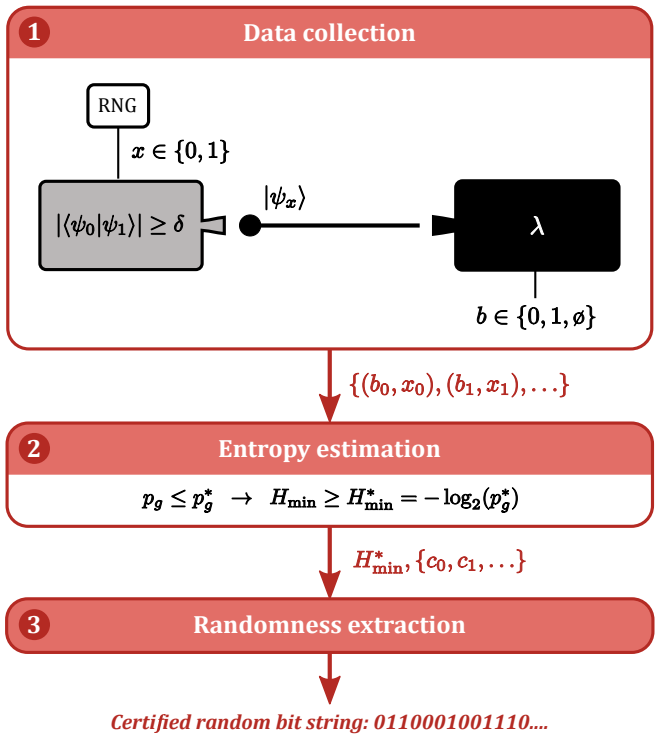}
  \caption{Steps of our QRNG protocol. \textbf{(1)} Data is generated in a prepare-and-measure setup. The prepared states are known to have a certain minimal overlap, hence the preparation device is a 'gray box', while nothing is assumed about the measurement device, which is a 'black box'. \textbf{(2)} From the collected data, a conditional probability distribution for outputs given inputs is estimated, and from this, a bound on the entropy in the output data is evaluated. \textbf{(3)} Based on the entropy bound, a string of certified perfectly random bits are extracted from the output data.}
  \label{fig.concept}
\end{figure}

\section{Protocol}

The conceptual scheme is illustrated in \figref{fig.concept}. The protocol consists in three steps. (1) data collection from measurements on quantum states, (2) estimation of the genuinely quantum entropy in the data, and (3) randomness extraction. 

In step (1), a preparation device takes a binary input $x \in \{0,1\}$ and emits a quantum system in state $\ket{\psi_x}$. The central assumption of the protocol is that the overlap of the two possible states is lower-bounded, $|\braket{\psi_0}{\psi_1}| \geqslant \delta$. In other words, we assume that the states are non-orthogonal and hence not deterministically distinguishable. However, a detailed description of the states is not required. For simplicity, we keep the states pure for now. At the end of this section, we discuss the precise assumptions which our protocol is based on.

The state is sent to a measurement device, which provides a ternary output $b \in \{0,1,\o \}$. The main idea of our protocol is that the measurement device performs USD. The goal is thus to maximize the probability of identifying which state has been prepared without errors, i.e. maximize $p(b=x)$ while ensuring that $p(b = \lnot x)=0$. While quantum theory allows for such a measurement, it imposes a minimal rate of inconclusive events $p(b= \o) \geqslant  \delta $ \cite{Barnett2009}. Note that this is a fundamental limit of quantum theory; if a better measurement were possible, this would have dramatic consequences, e.g.~instantaneous transmission of information. Importantly, it is not possible to predict in advance whether a particular round of the experiment will be conclusive or inconclusive. Clearly, if that were possible, then a better measurement could be implemented. Therefore, the occurrence of inconclusive events is a genuinely random quantum phenomena. 

The protocol exploits this randomness source in order to generate a final random bit string. In each round of the protocol, we thus define a bit $c$ which encodes whether this round was conclusive or not, i.e. $c = 0$ if $b=0,1$ and $c=1$ if $b=\o$. The value of $b$ when the measurement is conclusive (i.e. $b=0$ or $b=1$) will not be directly used for extracting randomness. This value is however important, and will be used in order to estimate the entropy in the data. One can understand this as verifying that the measurement device is indeed performing a USD measurement, i.e.~self-testing of the device.

Our goal is now to bound the amount of randomness in $c$ given the overlap $\delta$ and the observed $p(b|x)$, that is, the probability of obtaining output $b$ given preparation $x$. To see that the idea makes sense, consider first the ideal case in which the preparation device emits two states $\ket{\psi_x}$ with overlap $|\braket{\psi_0}{\psi_1}| = \delta$, and the measurement device implements a perfect USD. Here we have that $p(0|0) = p(1|1) = 1-\delta$, no errors $p(0|1)=p(1|0)=0$, and $p(\o|0)=p(\o|1)=\delta$. Hence the probability of guessing $c$ is $p_g = \delta$. In particular, for the choice $\delta=1/2$, a perfectly random bit can thus be certified. 

Now consider the general case, where the statistics are not assumed to originate from a perfect USD measurement, for instance due to unavoidable technical imperfections. Given the probabilities $p(b|x)$ and a bound on the overlap $\delta$, we show how to bound the probability $p_g$ of guessing $c$ for an observer with complete knowledge of the inner workings of the device, the input states, and the details of the measurement, which may vary from run to run. We label the measurement strategies by $\lambda$. The guessing probability averaged over inputs and measurement strategies, occurring with probabilities $p(x)$ and $p(\lambda)$ respectively, is then given by
\begin{equation}
\label{eq.pg}
p_g = \sum_x p(x) \sum_{\lambda} p(\lambda) \max\{ \Tr[\rho_x \Pi^\lambda_{\o} ] , 1-\Tr[\rho_x \Pi^\lambda_{\o} ] \} ,
\end{equation}
where $\rho_x = \ket{\psi_x}\bra{\psi_x}$, and $\Pi^\lambda_b$ are the elements of a three-outcome positive-operator-value measure (POVM) describing the measurement. To certify randomness, we need to upper bound $p_g$ over all possible measurement strategies which are consistent with the observed experimental data. Because the trace is invariant under unitary transformations, only the overlap of the input states matter, and not the states themselves. As we explain in \appref{sec.sdp}, upper bounds on $p_g$ can be established by means of semidefinite programming (SDP). Specifically,
\begin{equation}
\label{eq.pgbound}
p_g \leqslant p_g^* = \sum_{b,x} \nu_{bx} p(b|x)
\end{equation}
for any numbers $\nu_{bx}$ which fulfil that there exists four $2 \times 2$ hermitian matrices $H^{\lambda_0,\lambda_1}$, with $\lambda_0,\lambda_1=0,1$ such that
\begin{align}
\sum_x \rho_x & ( \frac{1}{2}\delta_{\lambda_x,0}\delta_{b,\o} + \frac{1}{2}\delta_{\lambda_x,1}(1-\delta_{b,\o}) - \nu_{bx} ) \nonumber \\
& + H^{\lambda_0,\lambda_1} - \frac{1}{2}\Tr[H^{\lambda_0,\lambda_1}]\mathbbm{1} \leqslant 0 .
\end{align}
Coefficients $\nu_{bx}$ that are optimal for particular data $p(b|x)$ can be found by SDP. However, given valid $\nu_{bx}$ and fixed $\delta$, the bound \eqref{eq.pgbound} holds for any $p(b|x)$. This implies that it is not necessary to run an SDP every time $p(b|x)$ is updated. One only needs to evaluate \eqref{eq.pgbound} which is a simple, linear function of the data, using fixed values of $\nu_{bx}$ (or a few tabulated values and take the tightest bound). This enables fast QRNG and simple incorporation of finite-size effects. Note that for perfect USD of states with overlap $\delta$, we find (numerically, using SDP to optimise $\nu_{bx}$), that our bound certifies $p_g \leqslant \delta$ \footnote{Note that this bound assumes $\delta \geq 1/2$. If $\delta < 1/2$, then the outcome will be most likely be conclusive, and therefore we have that $p_g \leqslant 1- \delta$.}. 

In step (2) of the protocol, from the experimental data of a number of runs, the input-output probability distribution $p(b|x)$ is estimated, and the bound \eqref{eq.pgbound} is evaluated. This also provides a bound on the genuinely quantum entropy in the string of raw bits $c$, given by the min-entropy 
\ba \label{eq.minentropy}
H_{\text{min}} = - \log_2 (p_g) .
\ea
The min-entropy quantifies the number of certified random bits that can be extracted per bit of the raw data \cite{Konig2009}. The final step (3) of the protocol consists in extracting a final random bit string via a randomness extraction procedure, based on the bound on $H_{\text{min}}$.

Finally, we discuss all assumptions required in our protocol. First, we assume that the input $x$ is generated independently from the devices, in particular $x$ should be independent from $\lambda$. In our experimental implementation, $x$ will be generated from a classical RNG (e.g. a pseudo randomness generator). The second assumptions concerns the overlap of the two prepared states. We assume that, in each round of the protocol, the two prepared states cannot be perfectly distinguished (using any possible quantum measurement procedure). If the two states are pure, it is possible to discriminate them without any error, at the price of having a minimal rate of inconclusive rounds, given by the overlap between the two states. Note that if the states are mixed, with overlapping support, then they cannot be distinguished unambiguously anymore. 
We assume that the two prepared states, $\rho_0$ and $\rho_1$ fulfill $F(\rho_0,\rho_1) \geqslant \delta$, where $F$ is the fidelity. This condition must hold with respect to any observer, in particular from the point of view of the measuring device. No additional information is available which allows picking out specific terms in any decomposition of the states. This ensures that $\rho_0$ and $\rho_1$ have a minimal indistinguishability from the point of view of the measuring device. Hence, no measurement procedure allowed in quantum theory would allow one to distinguish the two states better. In particular, no fault in the implementation of the measuring device can make the states more distinguishable. Since, without additional information, going from pure to mixed states with the same fidelity cannot help in distinguishing the states, taking $\rho_0$ and $\rho_1$ pure is the most conservative choice when bounding the guessing probability, and hence our bound above is general under this assumption. We note that our requirement is similar to assuming that the prepared states in different rounds are independent and identically distributed (i.i.d.) with respect to all observers, however it is strictly weaker as we do not need the states in every round to be the same, only that their relative fidelity is bounded \footnote{Note though, that the assumption cannot be relaxed to a bound on the average overlap over many rounds, as there is then a classical strategy reproducing the USD probability distribution.}. We also stress that there are no assumptions on the measurement device whatsoever.

\section{Implementations}

\begin{figure}[b!]
  \centering
  \includegraphics[width=0.95\columnwidth]{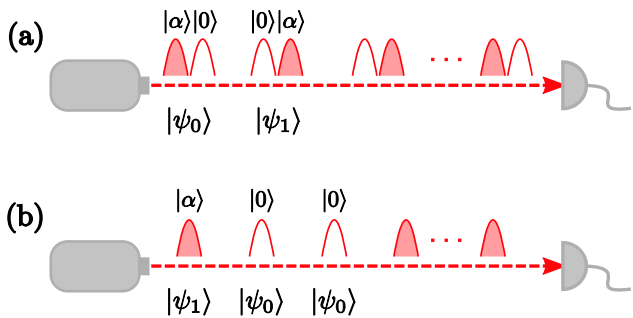}
  \caption{Implementations with weak coherent states encoded in time bins. \textbf{(a)} Two pulse scheme. For each pair of time bins, a laser emits a weak pulse in either the early or the late slot, corresponding to the states $\ket{\psi_0} = \ket{\alpha}\ket{0}$ and $\ket{\psi_1} = \ket{0}\ket{\alpha}$. Each bin is measured by a single-photon detector. If a click is registered in the early or late bin, the system outputs $b=0$ or $b=1$ respectively, while if no click is registered an inconclusive output is produced, $b=\o$. \textbf{(b)} Single-pulse scheme. The states are encoded in single pulses of weak coherent states or vacuum, $\ket{\psi_0} = \ket{0}$ and $\ket{\psi_1} = \ket{\alpha}$. When a click is registered, the output is $b=1$, while no click is treated as inconclusive $b=\o$. 
  }
  \label{fig.cowlike}
\end{figure}

We now discuss different possible implementations of our protocol. In the next section we report the experimental realization of two of these schemes, thus demonstrating practical relevance in situations involving loss and imperfections.

\emph{Implementation 1.} A first implementation uses a time-bin encoding, see \figref{fig.cowlike} (a). Here the two states are encoded by weak coherent pulses emitted in pairs of time-bins
\ba \label{time-bins}
\ket{\psi_0}=\ket{\alpha}\ket{0}  \quad , \quad \ket{\psi_1}=\ket{0}\ket{\alpha}. 
\ea
where $\ket{0}$ denotes the vacuum and $\ket{\alpha} = \exp(- \frac{|\alpha|^2}{2})  \sum_{n=0}^{\infty} \frac{\alpha^n}{n!}\ket{n}$ a coherent state with mean photon number $|\alpha|^2$. The overlap of these states is directly related to $|\alpha|^2$, namely
\ba
\label{eq_alphaoverlap}
\delta = |\braket{\psi_0}{\psi_1}| = \exp(-|\alpha|^2) . 
\ea
For weak pulses ($\alpha < 1$), the overlap is significant. Note that this encoding is reminiscent of the QKD protocol COW \cite{stucki2005}.

A practical advantage of this implementation is the simplicity of realizing the (optimal) USD measurement, which simply requires a single-photon detector with timing resolution sufficient to distinguish the two time bins. If a click is registered in the early (late) time-bin, the system outputs $b=0$ ($b=1$), while if no click is registered, the outcome is inconclusive $b=\o$. It is straightforward to check that in the absence of losses and noise, $p(b= \o) =  \exp(-|\alpha|^2)$, hence the measurement achieves the minimal rate of inconclusive outcomes, while giving no errors. 

In practice the measurement does not achieve the optimal USD exactly. Typically, detector inefficiency increases the inconclusive rate above that of the perfect USD, while detector dark counts increase the error rate. Nevertheless, randomness can still be extracted, as our protocol is sufficiently robust. 

\emph{Implementation 2.} Another possible implementation consists in using only a single weak coherent pulse, see \figref{fig.cowlike}~(b). The two non-orthogonal states are now simply 
\ba
\ket{\psi_0}=\ket{0}  \quad , \quad \ket{\psi_1}=\ket{\alpha}. 
\ea
This corresponds to an encoding in the photon number degree of freedom. The overlap between the two states is $ \delta = |\braket{\psi_0}{\psi_1}| = \exp(-\frac{|\alpha|^2}{2})$.  

As above, we use as a measurement a simple single-photon detector. If a click is registered, the output is $b=1$, while if no click is registered, the output is $b=\o$. The output $b=0$ thus never occurs. Note that the measurement is now effectively binary, and corresponds to a partial USD measurement, in the sense that it is only the state $\ket{\psi_1}$ that is identified unambiguously, hence $c=b$. So, the randomness is effectively generated from the state $\ket{\psi_1}$, while the state $\ket{\psi_0}$ is used to test that the device correctly performs the USD. Similarly to quantum key distribution protocols, it will then be advantageous to bias the input probability, i.e. setting $p(x=1)> p(x=0)$, in order to increase the output entropy. This will be discussed in the next section where we implement this protocol.

\emph{Further implementations.} Our approach can be implemented using more general encodings. For instance, a polarization encoding also represents a practical solution. Given two non-orthogonal states of polarization, the optimal USD measurement can be realized using a partial polarizer (i.e. polarization dependent losses) \cite{Huttner1996}. Encodings using frequency or spatial modes could also be considered.

\section{Experiments}

\begin{figure}
  \centering
  \includegraphics[width=0.95\columnwidth]{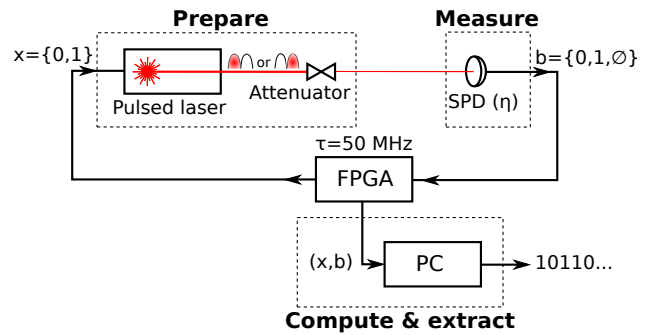}
  \caption{Experimental implementation of the QRNG. The preparation device corresponds here to the two pulse protocol.}
  \label{fig_exp}
\end{figure}

We have experimentally realized our QRNG based on USD, using the two main implementations discussed above, namely based on time-bins (two pulses) and photon number encodings (single pulse). Both implementations are essentially based on the same setup, with only minor modifications. 

\begin{figure*}[t!]
 \includegraphics[width=0.9\columnwidth]{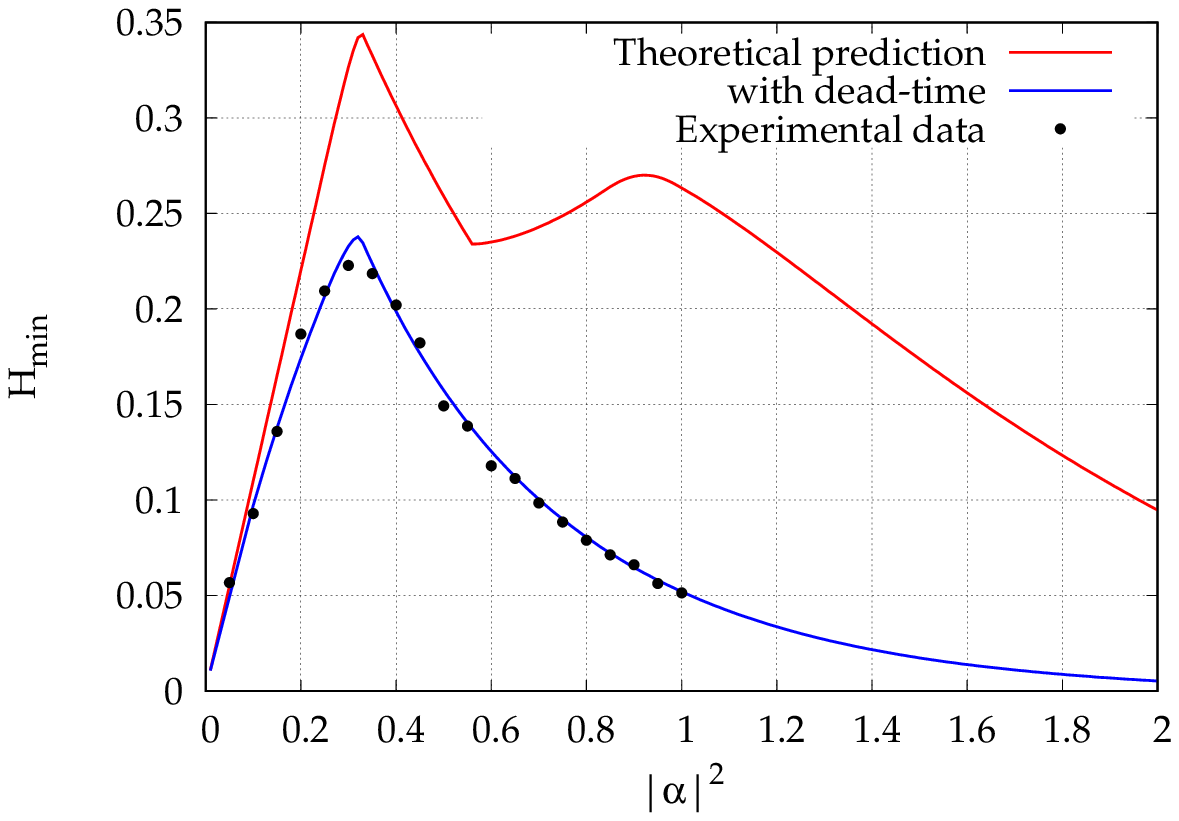}
 \includegraphics[width=0.9\columnwidth]{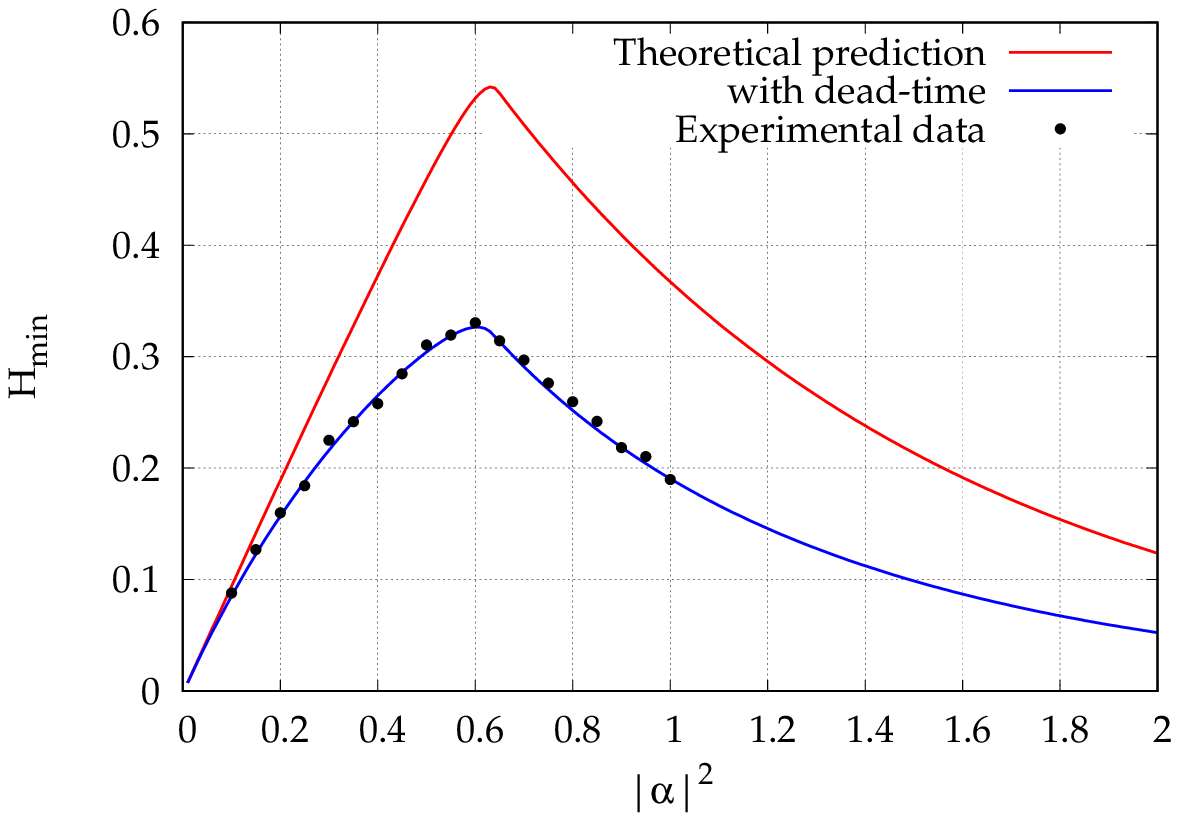}
  \caption{Min-entropy per raw bit generated by the QRNG as a function of the energy per pulse $|\alpha|^2$. The left plot represents the protocol with time-bin encoding (two pulses), while the right plot considers the photon number protocol (single pulse). In both plots, the red curve corresponds to the theoretical prediction obtained for a perfect single photon detector with an efficiency of 77\% (corresponding to our experimental value), but without considering dead-time. The blue curve considers the effect of detector dead-time, and shows good agreement with experimental data (black points).}
  \label{fig_hmin}
\end{figure*}

We first discuss the time-bin implementation. In order to generate the two non-orthogonal states \eqref{time-bins}, a field-programmable gate array (FPGA) triggers a fibered laser diode at a rate of 50\,MHz, as presented in \figref{fig_exp}. A pseudo-random generator generates the input $x$. If $x=1$ the electronic pulse is delayed by 10 ns, while nothing happens if $x=0$. This generates the states $\ket{\psi_1}$ and $\ket{\psi_0}$, respectively. At each trigger signal, the laser diode emits light pulses of 40 ps at 655\,nm. To set the appropriate light intensity, two adjustable attenuators are placed at the output of the laser after a 50/50 beam-splitter (BS). The second port of the BS is connected to a calibrated power meter which monitors the laser power, and the attenuation is adjusted based on this reading.

At the output of the source, the light is detected by a silicon avalanche photodiode single-photon detector (PerkinElmer - SPCM-AQR) with an efficiency of 77\% and a temporal jitter smaller than 1\,ns, which is enough to temporally discriminate the pulses separated by 10\,ns. The detector has around 300\,Hz of dark counts and a dead-time of 50\,ns. All the detection events are recorded by the FPGA. Every second, after taking data, the conditional probabilities $p(b|x)$ are evaluated. This generates 50\,M of raw bits, the entropy of which will be estimated via our protocol. The estimation of the probabilities $p(b|x)$ is made from a finite number of trials N. To take into account the error on the estimation of these probabilities due to finite statistics effect we use the Chernoff-Hoeffding tail inequality~\cite{hoeffding1963}, which provides an upper (lower) bound on the probability that the sum of random variables deviates from its expected value. From the experimental statistics $\xi(b|x) = \frac{n_{b,x}}{\sum_b n_{b,x}}$, where $n_{b,x}$ denote the number of events with outcome $b$ and input $x$, we get:
\begin{equation}
\xi(b|x) - t\left(\epsilon, \sum\nolimits_b n_{b,x}\right) \leqslant p(b|x)  \leqslant \xi(b|x) + t\left(\epsilon, \sum\nolimits_b n_{b,x}\right)
\end{equation}
with $t(\epsilon,N) := \sqrt{(\log(1/\epsilon)/(2N)}$. Here, $\epsilon$ is the confidence index, which represents the probability that the above relation is not satisfied. In our experiments, we choose $\epsilon= 10^{-9}$. From this, we can lower bound the relation of Eqs.~\eqref{eq.pgbound} by:
\begin{equation} \label{eq:pg_finitesize}
p^*_g\leqslant  p^N_g  = \sum_{b,x} \nu_{bx}\xi(b|x) + \sum_{b,x} | \nu_{bx} |Ê\, t\left(\epsilon,\sum\nolimits_b n_{b,x}\right).
\end{equation}
Note that the above bound is conservative, but essentially optimal when $t\left(\epsilon,\sum\nolimits_b n_{b,x}\right)$ is very small; a tighter bound can be obtained by further imposing that the distribution $p(b|x)$ is normalized. To generate the final bit string with quasi-perfect entropy, an extractor is applied to the raw bit string, with a compression factor which depends on the target entropy and the min-entropy contained in the raw data, $H_{\rm min} = -\log_2(p^N_g )$. Hence, the final bit rate of the QRNG is adapted in such a way that the min-entropy per output bit is constant. 

In our configuration, the light pulse energy is the only adjustable parameter that can be tuned to optimize the min-entropy per raw bit. \figref{fig_hmin} (left) represents the min-entropy as a function of $|\alpha|^2$ which is directly related to the overlap between the two states through \eqref{eq_alphaoverlap}. The upper red curve represents the theoretical prediction taking into account the finite statistic effect when we consider single-photon detection with an efficiency of 77\% (i.e.~matching our experimental value, but without saturation effects) \footnote{The two peaks of the top curve on the left plot in \figref{fig_hmin} arise from two different measurement strategies maximising the guessing probabilities, while reproducing the observed data, in different loss regimes. Disregarding finite size effects, for low loss, a USD strategy is optimal, and fixing the conclusive rate at 1/2 requires a choice of $|\alpha|^2 = \log(2)/\eta$. For increasing loss, a mixture of strategies where only one state is identified (i.e. one of the outputs 0 or 1 has vanishing probability) becomes optimal. In this case the inconclusive rate is $\delta^2$ and the optimal $|\alpha|^2 = \log(\sqrt{2})$.}. The dead-time effect can be modeled by applying the correction factor $c_d = \frac{N_{\rm det}}{1+t_d^* N_{\rm det}}$ on the detection probabilities, where $N_{\rm det}$ and $t_d^*$ correspond to the total number of detection and the effective dead-time of the detector, respectively. This model is usually employed with a uniform and continuous source of photons and $t_d$ will correspond to the detector dead-time. In a pulsed regime, we can use the same model with an effective dead-time which depends on the dead-time of the detector and the repetition rate of the laser. In our configuration, we estimate that $t_d^*$ is equal to 34\,ns. Taking experimental imperfections into account, we see that a maximal entropy of 0.22 is obtained for a pulse energy of $|\alpha|^2 = 0.3$, which allows us to generate 11\.MHz of final random bits after extraction. Here the error rate is typically around $4\times10^{-4}$.

Let us now move to the second implementation, using photon number as a degree of freedom. In this single pulse approach, the only difference is the configuration of the FPGA. Indeed, instead of delaying or not the optical pulse, the FPGA now sends or not the pulse (hence the emitted state is the vacuum) with a probability $p(x=0) = 1/8$. This probability bias is optimal when we consider a block size of 50 Mbits. Note that the bias can be increased for a larger block size, in order to increase the generation rate. As shown in \figref{fig_hmin} (right), we obtain here an entropy per bit of 0.33 for $|\alpha|^2 = 0.60$, which allows us to generate 16.5\,MHz of final random bits after extraction. 

\begin{figure}
\includegraphics[width = \columnwidth]{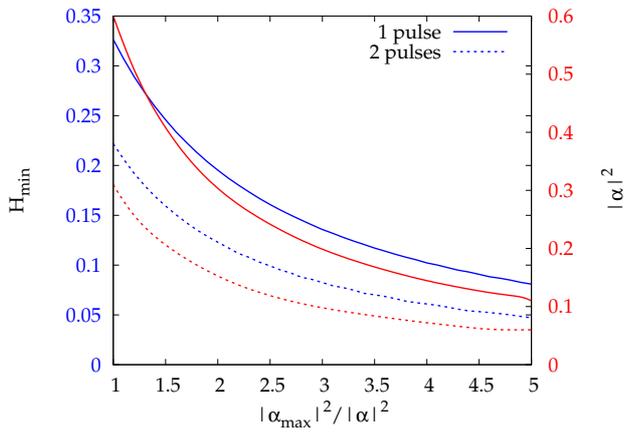}
\caption{\label{fig_delta} Bound on the entropy when taking possible power fluctuations of the source into account. The min-entropy is plotted as a function of the ratio $ |\alpha_{\max} |^2 /  |\alpha |^2$, where $ |\alpha_{\max} |^2$ is the maximal energy per pulse and $|\alpha |^2$ is  the average pulse energy of the source. For both protocols, we plot the min-entropy in blue and the corresponding average energy per pulse of the source in red. Note that even when the maximal pulse energy is assumed to be as high as five times larger than the mean energy, a reasonable amount of entropy can still be certified.}
\end{figure}

Finally, let us comment on the justification of the assumptions required in our protocol. These are essentially the same in both configurations. The first assumption concerns the fact that the generation of the input $x$ must be independent from the devices. This is easily realized since $x$ is generated by the FPGA. The second assumption is the crucial one. Here we must ensure that the pulse energy of the source is well characterized, in order to satisfy the assumption that the overlap of the two states is at least $\delta$. Importantly, the overlap must be bounded in each round of the protocol, which can be delicate if the source features non-negligible power fluctuations, e.g.~due to instabilities in the laser itself or in the attenuator in \figref{fig_exp}. When the energy per pulse becomes higher, the overlap of the output states decreases, hence if such fluctuations are not accounted for, the overlap may decrease below $\delta$, violating the assumption. There are several possibilities to address this point. First, one can choose $\delta$ in a conservative manner, and not based directly on the (mean) power of the source $|\alpha|^2$, but rather with respect to a maximal energy per pulse $|\alpha_{\max}|^2$. That is, the protocol can be run under the assumptions of a given overlap $\delta $ (corresponding to $|\alpha_{\max}|^2$), while the mean pulse energy of the source $|\alpha|^2$ corresponds in fact to a much larger overlap, i.e. $|\braket{\psi_0}{\psi_1}| \gg \delta$. This will decrease the entropy per bit, as shown in \figref{fig_delta}, but final randomness can nevertheless still certified, given that power fluctuations are not too large. Another possibility would be to use an optical fuse~\cite{Todoroki2004}, i.e. an optical channel breaking down above a certain threshold intensity. 

\section{Conclusion}

We have proposed an approach to quantum random number generation based on USD measurements. The protocol is in prepare-and-measure configuration, and based on the fact that the occurrence of inconclusive events in unambiguous state discrimination must be genuinely random. Our protocol offers semi-DI security, in the sense that the amount of trust in the physical implementation is low. Specifically, the main assumption is a bound on the overlap of the prepared states, but no assumption about the measurement device is needed. At the same time, the protocol is practical, which we demonstrated by implementing it using a simple optical setup. We achieved a random bit rate of 16.5\.MHz, which is comparable to commercial QRNGs~\cite{QRNG}. Our approach thus combines strong security, allowing the user to monitor the entropy of the output in real time, as well as ease of implementation and high rates.

\emph{Note added.} The setup of the single-pulse protocol was independently discussed by the authors of Ref. \cite{Himbeck}, but analyzed under different technical and security assumptions.

\emph{Acknowledgements.} We thank Stefano Pironio and Eric Woodhead for discussions. We acknowledge financial support from the Swiss National Science Foundation (Starting grant DIAQ, Grant SNF $200021_159592$, and QSIT) and the AXA Chair in Quantum Information Science, a Severo Ochoa Grant SEV-2015-0522 and Fundacion CELLEX.

\bibliography{USDrandomness}

\begin{thebibliography}{53}%
\makeatletter
\providecommand \@ifxundefined [1]{%
 \@ifx{#1\undefined}
}%
\providecommand \@ifnum [1]{%
 \ifnum #1\expandafter \@firstoftwo
 \else \expandafter \@secondoftwo
 \fi
}%
\providecommand \@ifx [1]{%
 \ifx #1\expandafter \@firstoftwo
 \else \expandafter \@secondoftwo
 \fi
}%
\providecommand \natexlab [1]{#1}%
\providecommand \enquote  [1]{``#1''}%
\providecommand \bibnamefont  [1]{#1}%
\providecommand \bibfnamefont [1]{#1}%
\providecommand \citenamefont [1]{#1}%
\providecommand \href@noop [0]{\@secondoftwo}%
\providecommand \href [0]{\begingroup \@sanitize@url \@href}%
\providecommand \@href[1]{\@@startlink{#1}\@@href}%
\providecommand \@@href[1]{\endgroup#1\@@endlink}%
\providecommand \@sanitize@url [0]{\catcode `\\12\catcode `\$12\catcode
  `\&12\catcode `\#12\catcode `\^12\catcode `\_12\catcode `\%12\relax}%
\providecommand \@@startlink[1]{}%
\providecommand \@@endlink[0]{}%
\providecommand \url  [0]{\begingroup\@sanitize@url \@url }%
\providecommand \@url [1]{\endgroup\@href {#1}{\urlprefix }}%
\providecommand \urlprefix  [0]{URL }%
\providecommand \Eprint [0]{\href }%
\providecommand \doibase [0]{http://dx.doi.org/}%
\providecommand \selectlanguage [0]{\@gobble}%
\providecommand \bibinfo  [0]{\@secondoftwo}%
\providecommand \bibfield  [0]{\@secondoftwo}%
\providecommand \translation [1]{[#1]}%
\providecommand \BibitemOpen [0]{}%
\providecommand \bibitemStop [0]{}%
\providecommand \bibitemNoStop [0]{.\EOS\space}%
\providecommand \EOS [0]{\spacefactor3000\relax}%
\providecommand \BibitemShut  [1]{\csname bibitem#1\endcsname}%
\let\auto@bib@innerbib\@empty
\bibitem [{\citenamefont {Hayes}(2001)}]{Hayes2001}%
  \BibitemOpen
  \bibfield  {author} {\bibinfo {author} {\bibfnamefont {B.}~\bibnamefont
  {Hayes}},\ }\bibfield  {title} {\enquote {\bibinfo {title} {{Randomness as a
  Resource}},}\ }\href {\doibase 10.1511/2001.4.300} {\bibfield  {journal}
  {\bibinfo  {journal} {Am. Sci.}\ }\textbf {\bibinfo {volume} {89}},\ \bibinfo
  {pages} {300} (\bibinfo {year} {2001})}\BibitemShut {NoStop}%
\bibitem [{\citenamefont {Herrero-Collantes}\ and\ \citenamefont
  {Garcia-Escartin}(2017)}]{Herrero-collantes2014}%
  \BibitemOpen
  \bibfield  {author} {\bibinfo {author} {\bibfnamefont {M.}~\bibnamefont
  {Herrero-Collantes}}\ and\ \bibinfo {author} {\bibfnamefont {J.~C.}\
  \bibnamefont {Garcia-Escartin}},\ }\bibfield  {title} {\enquote {\bibinfo
  {title} {Quantum random number generators},}\ }\href {\doibase
  10.1103/RevModPhys.89.015004} {\bibfield  {journal} {\bibinfo  {journal}
  {Rev. Mod. Phys.}\ }\textbf {\bibinfo {volume} {89}},\ \bibinfo {pages}
  {015004} (\bibinfo {year} {2017})}\BibitemShut {NoStop}%
\bibitem [{\citenamefont {Bera}\ \emph {et~al.}(2016)\citenamefont {Bera},
  \citenamefont {Acin}, \citenamefont {Kus}, \citenamefont {Mitchell},\ and\
  \citenamefont {Lewenstein}}]{Bera2016}%
  \BibitemOpen
  \bibfield  {author} {\bibinfo {author} {\bibfnamefont {M.N.}\ \bibnamefont
  {Bera}}, \bibinfo {author} {\bibfnamefont {A.}~\bibnamefont {Acin}}, \bibinfo
  {author} {\bibfnamefont {M.}~\bibnamefont {Kus}}, \bibinfo {author}
  {\bibfnamefont {M.}~\bibnamefont {Mitchell}}, \ and\ \bibinfo {author}
  {\bibfnamefont {M.}~\bibnamefont {Lewenstein}},\ }\bibfield  {title}
  {\enquote {\bibinfo {title} {{Randomness in Quantum Mechanics: Philosophy,
  Physics and Technology}},}\ }\href {https://arxiv.org/abs/1611.02176}
  {\bibfield  {journal} {\bibinfo  {journal} {arXiv:1611.02176 [quant-ph]}\ }
  (\bibinfo {year} {2016})}\BibitemShut {NoStop}%
\bibitem [{\citenamefont {Rarity}\ \emph {et~al.}(1994)\citenamefont {Rarity},
  \citenamefont {Owens},\ and\ \citenamefont {Tapster}}]{rarity1994}%
  \BibitemOpen
  \bibfield  {author} {\bibinfo {author} {\bibfnamefont {J.G.}\ \bibnamefont
  {Rarity}}, \bibinfo {author} {\bibfnamefont {P.C.M.}\ \bibnamefont {Owens}},
  \ and\ \bibinfo {author} {\bibfnamefont {P.R.}\ \bibnamefont {Tapster}},\
  }\bibfield  {title} {\enquote {\bibinfo {title} {Quantum random-number
  generation and key sharing},}\ }\href {\doibase 10.1080/09500349414552281}
  {\bibfield  {journal} {\bibinfo  {journal} {J. Mod. Opt.}\ }\textbf {\bibinfo
  {volume} {41}},\ \bibinfo {pages} {2435--2444} (\bibinfo {year}
  {1994})}\BibitemShut {NoStop}%
\bibitem [{\citenamefont {Stefanov}\ \emph {et~al.}(2000)\citenamefont
  {Stefanov}, \citenamefont {Gisin}, \citenamefont {Guinnard}, \citenamefont
  {Guinnard},\ and\ \citenamefont {Zbinden}}]{stefanov2000}%
  \BibitemOpen
  \bibfield  {author} {\bibinfo {author} {\bibfnamefont {A.}~\bibnamefont
  {Stefanov}}, \bibinfo {author} {\bibfnamefont {N.}~\bibnamefont {Gisin}},
  \bibinfo {author} {\bibfnamefont {O.}~\bibnamefont {Guinnard}}, \bibinfo
  {author} {\bibfnamefont {L.}~\bibnamefont {Guinnard}}, \ and\ \bibinfo
  {author} {\bibfnamefont {H.}~\bibnamefont {Zbinden}},\ }\bibfield  {title}
  {\enquote {\bibinfo {title} {Optical quantum random number generator},}\
  }\href {https://doi.org/10.1080/09500340008233380} {\bibfield  {journal}
  {\bibinfo  {journal} {J. Mod. Opt.}\ }\textbf {\bibinfo {volume} {47}},\
  \bibinfo {pages} {595--598} (\bibinfo {year} {2000})}\BibitemShut {NoStop}%
\bibitem [{\citenamefont {Jennewein}\ \emph {et~al.}(2000)\citenamefont
  {Jennewein}, \citenamefont {Achleitner}, \citenamefont {Weihs}, \citenamefont
  {Weinfurter},\ and\ \citenamefont {Zeilinger}}]{jennewein2000}%
  \BibitemOpen
  \bibfield  {author} {\bibinfo {author} {\bibfnamefont {T.}~\bibnamefont
  {Jennewein}}, \bibinfo {author} {\bibfnamefont {U.}~\bibnamefont
  {Achleitner}}, \bibinfo {author} {\bibfnamefont {G.}~\bibnamefont {Weihs}},
  \bibinfo {author} {\bibfnamefont {H.}~\bibnamefont {Weinfurter}}, \ and\
  \bibinfo {author} {\bibfnamefont {A.}~\bibnamefont {Zeilinger}},\ }\bibfield
  {title} {\enquote {\bibinfo {title} {A fast and compact quantum random number
  generator},}\ }\href {\doibase http://dx.doi.org/10.1063/1.1150518}
  {\bibfield  {journal} {\bibinfo  {journal} {Rev. Sci. Instrum.}\ }\textbf
  {\bibinfo {volume} {71}},\ \bibinfo {pages} {1675--1680} (\bibinfo {year}
  {2000})}\BibitemShut {NoStop}%
\bibitem [{\citenamefont {Dynes}\ \emph {et~al.}(2008)\citenamefont {Dynes},
  \citenamefont {Yuan}, \citenamefont {Sharpe},\ and\ \citenamefont
  {Shields}}]{dynes2008}%
  \BibitemOpen
  \bibfield  {author} {\bibinfo {author} {\bibfnamefont {J.~F.}\ \bibnamefont
  {Dynes}}, \bibinfo {author} {\bibfnamefont {Z.~L.}\ \bibnamefont {Yuan}},
  \bibinfo {author} {\bibfnamefont {A.~W.}\ \bibnamefont {Sharpe}}, \ and\
  \bibinfo {author} {\bibfnamefont {A.~J.}\ \bibnamefont {Shields}},\
  }\bibfield  {title} {\enquote {\bibinfo {title} {A high speed, postprocessing
  free, quantum random number generator},}\ }\href {\doibase
  http://dx.doi.org/10.1063/1.2961000} {\bibfield  {journal} {\bibinfo
  {journal} {Appl. Phys. Lett.}\ }\textbf {\bibinfo {volume} {93}},\ \bibinfo
  {eid} {031109} (\bibinfo {year} {2008})}\BibitemShut {NoStop}%
\bibitem [{\citenamefont {Wahl}\ \emph {et~al.}(2011)\citenamefont {Wahl},
  \citenamefont {Leifgen}, \citenamefont {Berlin}, \citenamefont
  {R{\"o}hlicke}, \citenamefont {Rahn},\ and\ \citenamefont
  {Benson}}]{wahl2011}%
  \BibitemOpen
  \bibfield  {author} {\bibinfo {author} {\bibfnamefont {M.}~\bibnamefont
  {Wahl}}, \bibinfo {author} {\bibfnamefont {M.}~\bibnamefont {Leifgen}},
  \bibinfo {author} {\bibfnamefont {M.}~\bibnamefont {Berlin}}, \bibinfo
  {author} {\bibfnamefont {T.}~\bibnamefont {R{\"o}hlicke}}, \bibinfo {author}
  {\bibfnamefont {H.-J.}\ \bibnamefont {Rahn}}, \ and\ \bibinfo {author}
  {\bibfnamefont {O.}~\bibnamefont {Benson}},\ }\bibfield  {title} {\enquote
  {\bibinfo {title} {An ultrafast quantum random number generator with provably
  bounded output bias based on photon arrival time measurements},}\ }\href
  {\doibase http://dx.doi.org/10.1063/1.3578456} {\bibfield  {journal}
  {\bibinfo  {journal} {Appl. Phys. Lett.}\ }\textbf {\bibinfo {volume} {98}},\
  \bibinfo {eid} {171105} (\bibinfo {year} {2011})}\BibitemShut {NoStop}%
\bibitem [{\citenamefont {Nie}\ \emph {et~al.}(2014)\citenamefont {Nie},
  \citenamefont {Zhang}, \citenamefont {Zhang}, \citenamefont {Wang},
  \citenamefont {Ma}, \citenamefont {Zhang},\ and\ \citenamefont
  {Pan}}]{nie2014}%
  \BibitemOpen
  \bibfield  {author} {\bibinfo {author} {\bibfnamefont {You-Qi}\ \bibnamefont
  {Nie}}, \bibinfo {author} {\bibfnamefont {Hong-Fei}\ \bibnamefont {Zhang}},
  \bibinfo {author} {\bibfnamefont {Zhen}\ \bibnamefont {Zhang}}, \bibinfo
  {author} {\bibfnamefont {Jian}\ \bibnamefont {Wang}}, \bibinfo {author}
  {\bibfnamefont {Xiongfeng}\ \bibnamefont {Ma}}, \bibinfo {author}
  {\bibfnamefont {Jun}\ \bibnamefont {Zhang}}, \ and\ \bibinfo {author}
  {\bibfnamefont {Jian-Wei}\ \bibnamefont {Pan}},\ }\bibfield  {title}
  {\enquote {\bibinfo {title} {Practical and fast quantum random number
  generation based on photon arrival time relative to external reference},}\
  }\href {\doibase http://dx.doi.org/10.1063/1.4863224} {\bibfield  {journal}
  {\bibinfo  {journal} {Appl. Phys. Lett.}\ }\textbf {\bibinfo {volume}
  {104}},\ \bibinfo {eid} {051110} (\bibinfo {year} {2014})}\BibitemShut
  {NoStop}%
\bibitem [{\citenamefont {Stip\v{c}evi\'c}\ and\ \citenamefont
  {Rogina}(2007)}]{stipcevic2007}%
  \BibitemOpen
  \bibfield  {author} {\bibinfo {author} {\bibfnamefont {M.}~\bibnamefont
  {Stip\v{c}evi\'c}}\ and\ \bibinfo {author} {\bibfnamefont {B.~Medved}\
  \bibnamefont {Rogina}},\ }\bibfield  {title} {\enquote {\bibinfo {title}
  {Quantum random number generator based on photonic emission in
  semiconductors},}\ }\href {\doibase http://dx.doi.org/10.1063/1.2720728}
  {\bibfield  {journal} {\bibinfo  {journal} {Rev. Sci. Instrum.}\ }\textbf
  {\bibinfo {volume} {78}},\ \bibinfo {eid} {045104} (\bibinfo {year}
  {2007})}\BibitemShut {NoStop}%
\bibitem [{\citenamefont {Stip{\v{c}}evi{\'{c}}}\ and\ \citenamefont
  {Ursin}(2015)}]{Stipcevic2015}%
  \BibitemOpen
  \bibfield  {author} {\bibinfo {author} {\bibfnamefont {M.}~\bibnamefont
  {Stip{\v{c}}evi{\'{c}}}}\ and\ \bibinfo {author} {\bibfnamefont
  {R.}~\bibnamefont {Ursin}},\ }\bibfield  {title} {\enquote {\bibinfo {title}
  {{An On-Demand Optical Quantum Random Number Generator with In-Future Action
  and Ultra-Fast Response}},}\ }\href {\doibase 10.1038/srep10214} {\bibfield
  {journal} {\bibinfo  {journal} {Sci. Rep.}\ }\textbf {\bibinfo {volume}
  {5}},\ \bibinfo {pages} {10214} (\bibinfo {year} {2015})}\BibitemShut
  {NoStop}%
\bibitem [{\citenamefont {Qi}\ \emph {et~al.}(2010)\citenamefont {Qi},
  \citenamefont {Chi}, \citenamefont {Lo},\ and\ \citenamefont
  {Qian}}]{qi2010}%
  \BibitemOpen
  \bibfield  {author} {\bibinfo {author} {\bibfnamefont {Bing}\ \bibnamefont
  {Qi}}, \bibinfo {author} {\bibfnamefont {Yue-Meng}\ \bibnamefont {Chi}},
  \bibinfo {author} {\bibfnamefont {Hoi-Kwong}\ \bibnamefont {Lo}}, \ and\
  \bibinfo {author} {\bibfnamefont {Li}~\bibnamefont {Qian}},\ }\bibfield
  {title} {\enquote {\bibinfo {title} {High-speed quantum random number
  generation by measuring phase noise of a single-mode laser},}\ }\href
  {\doibase 10.1364/OL.35.000312} {\bibfield  {journal} {\bibinfo  {journal}
  {Opt. Lett.}\ }\textbf {\bibinfo {volume} {35}},\ \bibinfo {pages} {312--314}
  (\bibinfo {year} {2010})}\BibitemShut {NoStop}%
\bibitem [{\citenamefont {Uchida}\ \emph {et~al.}(2008)\citenamefont {Uchida},
  \citenamefont {Amano}, \citenamefont {Inoue}, \citenamefont {Hirano},
  \citenamefont {Naito}, \citenamefont {Someya}, \citenamefont {OowadaIsao},
  \citenamefont {Kurashige}, \citenamefont {Shiki}, \citenamefont {Yoshimori},
  \citenamefont {Yoshimura},\ and\ \citenamefont {Davis}}]{uchida2008}%
  \BibitemOpen
  \bibfield  {author} {\bibinfo {author} {\bibfnamefont {A.}~\bibnamefont
  {Uchida}}, \bibinfo {author} {\bibfnamefont {K.}~\bibnamefont {Amano}},
  \bibinfo {author} {\bibfnamefont {M.}~\bibnamefont {Inoue}}, \bibinfo
  {author} {\bibfnamefont {K.}~\bibnamefont {Hirano}}, \bibinfo {author}
  {\bibfnamefont {S.}~\bibnamefont {Naito}}, \bibinfo {author} {\bibfnamefont
  {H.}~\bibnamefont {Someya}}, \bibinfo {author} {\bibnamefont {OowadaIsao}},
  \bibinfo {author} {\bibfnamefont {T.}~\bibnamefont {Kurashige}}, \bibinfo
  {author} {\bibfnamefont {M.}~\bibnamefont {Shiki}}, \bibinfo {author}
  {\bibfnamefont {S.}~\bibnamefont {Yoshimori}}, \bibinfo {author}
  {\bibfnamefont {K.}~\bibnamefont {Yoshimura}}, \ and\ \bibinfo {author}
  {\bibfnamefont {P.}~\bibnamefont {Davis}},\ }\bibfield  {title} {\enquote
  {\bibinfo {title} {Fast physical random bit generation with chaotic
  semiconductor lasers},}\ }\href {http://dx.doi.org/10.1038/nphoton.2008.227}
  {\bibfield  {journal} {\bibinfo  {journal} {Nat. Photon.}\ }\textbf {\bibinfo
  {volume} {2}},\ \bibinfo {pages} {728--732} (\bibinfo {year}
  {2008})}\BibitemShut {NoStop}%
\bibitem [{\citenamefont {Abell\'{a}n}\ \emph {et~al.}(2014)\citenamefont
  {Abell\'{a}n}, \citenamefont {Amaya}, \citenamefont {Jofre}, \citenamefont
  {Curty}, \citenamefont {Ac\'{i}n}, \citenamefont {Capmany}, \citenamefont
  {Pruneri},\ and\ \citenamefont {Mitchell}}]{abellan2014}%
  \BibitemOpen
  \bibfield  {author} {\bibinfo {author} {\bibfnamefont {C.}~\bibnamefont
  {Abell\'{a}n}}, \bibinfo {author} {\bibfnamefont {W.}~\bibnamefont {Amaya}},
  \bibinfo {author} {\bibfnamefont {M.}~\bibnamefont {Jofre}}, \bibinfo
  {author} {\bibfnamefont {M.}~\bibnamefont {Curty}}, \bibinfo {author}
  {\bibfnamefont {A.}~\bibnamefont {Ac\'{i}n}}, \bibinfo {author}
  {\bibfnamefont {J.}~\bibnamefont {Capmany}}, \bibinfo {author} {\bibfnamefont
  {V.}~\bibnamefont {Pruneri}}, \ and\ \bibinfo {author} {\bibfnamefont
  {M.~W.}\ \bibnamefont {Mitchell}},\ }\bibfield  {title} {\enquote {\bibinfo
  {title} {Ultra-fast quantum randomness generation by accelerated phase
  diffusion in a pulsed laser diode},}\ }\href {\doibase 10.1364/OE.22.001645}
  {\bibfield  {journal} {\bibinfo  {journal} {Opt. Express}\ }\textbf {\bibinfo
  {volume} {22}},\ \bibinfo {pages} {1645--1654} (\bibinfo {year}
  {2014})}\BibitemShut {NoStop}%
\bibitem [{\citenamefont {Gabriel}\ \emph {et~al.}(2010)\citenamefont
  {Gabriel}, \citenamefont {Wittmann}, \citenamefont {Sych}, \citenamefont
  {Dong}, \citenamefont {Mauerer}, \citenamefont {Andersen}, \citenamefont
  {Marquardt},\ and\ \citenamefont {Leuchs}}]{gabriel2010}%
  \BibitemOpen
  \bibfield  {author} {\bibinfo {author} {\bibfnamefont {C.}~\bibnamefont
  {Gabriel}}, \bibinfo {author} {\bibfnamefont {C.}~\bibnamefont {Wittmann}},
  \bibinfo {author} {\bibfnamefont {D.}~\bibnamefont {Sych}}, \bibinfo {author}
  {\bibfnamefont {R.}~\bibnamefont {Dong}}, \bibinfo {author} {\bibfnamefont
  {W.}~\bibnamefont {Mauerer}}, \bibinfo {author} {\bibfnamefont {U.~L.}\
  \bibnamefont {Andersen}}, \bibinfo {author} {\bibfnamefont {C.}~\bibnamefont
  {Marquardt}}, \ and\ \bibinfo {author} {\bibfnamefont {G.}~\bibnamefont
  {Leuchs}},\ }\bibfield  {title} {\enquote {\bibinfo {title} {A generator for
  unique quantum random numbers based on vacuum states},}\ }\href@noop {}
  {\bibfield  {journal} {\bibinfo  {journal} {Nat. Photon.}\ }\textbf {\bibinfo
  {volume} {4}},\ \bibinfo {pages} {711--715} (\bibinfo {year}
  {2010})}\BibitemShut {NoStop}%
\bibitem [{\citenamefont {Symul}\ \emph {et~al.}(2011)\citenamefont {Symul},
  \citenamefont {Assad},\ and\ \citenamefont {Lam}}]{symul2011}%
  \BibitemOpen
  \bibfield  {author} {\bibinfo {author} {\bibfnamefont {T.}~\bibnamefont
  {Symul}}, \bibinfo {author} {\bibfnamefont {S.~M.}\ \bibnamefont {Assad}}, \
  and\ \bibinfo {author} {\bibfnamefont {P.~K.}\ \bibnamefont {Lam}},\
  }\bibfield  {title} {\enquote {\bibinfo {title} {Real time demonstration of
  high bitrate quantum random number generation with coherent laser light},}\
  }\href {\doibase http://dx.doi.org/10.1063/1.3597793} {\bibfield  {journal}
  {\bibinfo  {journal} {Appl. Phys. Lett.}\ }\textbf {\bibinfo {volume} {98}},\
  \bibinfo {eid} {231103} (\bibinfo {year} {2011})}\BibitemShut {NoStop}%
\bibitem [{\citenamefont {Sanguinetti}\ \emph {et~al.}(2014)\citenamefont
  {Sanguinetti}, \citenamefont {Martin}, \citenamefont {Zbinden},\ and\
  \citenamefont {Gisin}}]{sanguinetti2014}%
  \BibitemOpen
  \bibfield  {author} {\bibinfo {author} {\bibfnamefont {B.}~\bibnamefont
  {Sanguinetti}}, \bibinfo {author} {\bibfnamefont {A.}~\bibnamefont {Martin}},
  \bibinfo {author} {\bibfnamefont {H.}~\bibnamefont {Zbinden}}, \ and\
  \bibinfo {author} {\bibfnamefont {N.}~\bibnamefont {Gisin}},\ }\bibfield
  {title} {\enquote {\bibinfo {title} {Quantum random number generation on a
  mobile phone},}\ }\href {\doibase 10.1103/PhysRevX.4.031056} {\bibfield
  {journal} {\bibinfo  {journal} {Phys. Rev. X}\ }\textbf {\bibinfo {volume}
  {4}},\ \bibinfo {pages} {031056} (\bibinfo {year} {2014})}\BibitemShut
  {NoStop}%
\bibitem [{\citenamefont {Nisan}\ and\ \citenamefont
  {Ta-Shma}(1999)}]{nisan1999}%
  \BibitemOpen
  \bibfield  {author} {\bibinfo {author} {\bibfnamefont {Noam}\ \bibnamefont
  {Nisan}}\ and\ \bibinfo {author} {\bibfnamefont {Amnon}\ \bibnamefont
  {Ta-Shma}},\ }\bibfield  {title} {\enquote {\bibinfo {title} {Extracting
  randomness: A survey and new constructions},}\ }\href {\doibase
  http://dx.doi.org/10.1006/jcss.1997.1546} {\bibfield  {journal} {\bibinfo
  {journal} {J. Comput. Syst. Sci. Int.}\ }\textbf {\bibinfo {volume} {58}},\
  \bibinfo {pages} {148 -- 173} (\bibinfo {year} {1999})}\BibitemShut {NoStop}%
\bibitem [{\citenamefont {Frauchiger}\ \emph {et~al.}(2013)\citenamefont
  {Frauchiger}, \citenamefont {Renner},\ and\ \citenamefont
  {Troyer}}]{frauchiger2013}%
  \BibitemOpen
  \bibfield  {author} {\bibinfo {author} {\bibfnamefont {Daniela}\ \bibnamefont
  {Frauchiger}}, \bibinfo {author} {\bibfnamefont {Renato}\ \bibnamefont
  {Renner}}, \ and\ \bibinfo {author} {\bibfnamefont {Matthias}\ \bibnamefont
  {Troyer}},\ }\bibfield  {title} {\enquote {\bibinfo {title} {True randomness
  from realistic quantum devices},}\ }\href@noop {} {\bibfield  {journal}
  {\bibinfo  {journal} {arXiv preprint arXiv:1311.4547}\ } (\bibinfo {year}
  {2013})}\BibitemShut {NoStop}%
\bibitem [{\citenamefont {Ma}\ \emph {et~al.}(2013)\citenamefont {Ma},
  \citenamefont {Xu}, \citenamefont {Xu}, \citenamefont {Tan}, \citenamefont
  {Qi},\ and\ \citenamefont {Lo}}]{ma2013}%
  \BibitemOpen
  \bibfield  {author} {\bibinfo {author} {\bibfnamefont {Xiongfeng}\
  \bibnamefont {Ma}}, \bibinfo {author} {\bibfnamefont {Feihu}\ \bibnamefont
  {Xu}}, \bibinfo {author} {\bibfnamefont {He}~\bibnamefont {Xu}}, \bibinfo
  {author} {\bibfnamefont {Xiaoqing}\ \bibnamefont {Tan}}, \bibinfo {author}
  {\bibfnamefont {Bing}\ \bibnamefont {Qi}}, \ and\ \bibinfo {author}
  {\bibfnamefont {Hoi-Kwong}\ \bibnamefont {Lo}},\ }\bibfield  {title}
  {\enquote {\bibinfo {title} {Postprocessing for quantum random-number
  generators: Entropy evaluation and randomness extraction},}\ }\href {\doibase
  10.1103/PhysRevA.87.062327} {\bibfield  {journal} {\bibinfo  {journal} {Phys.
  Rev. A}\ }\textbf {\bibinfo {volume} {87}},\ \bibinfo {pages} {062327}
  (\bibinfo {year} {2013})}\BibitemShut {NoStop}%
\bibitem [{\citenamefont {Colbeck}(2009)}]{colbeckPhD}%
  \BibitemOpen
  \bibfield  {author} {\bibinfo {author} {\bibfnamefont {R.}~\bibnamefont
  {Colbeck}},\ }\href {https://arxiv.org/abs/0911.3814} {\enquote {\bibinfo
  {title} {Quantum and relativistic protocols for secure multi-party
  computation},}\ }\bibinfo {howpublished} {Ph.D. Thesis, University of
  Cambridge} (\bibinfo {year} {2009}),\ \bibinfo {note} {arXiv:0911.3814
  [quant-ph]}\BibitemShut {NoStop}%
\bibitem [{\citenamefont {Pironio}\ \emph {et~al.}(2010)\citenamefont
  {Pironio}, \citenamefont {Ac\'in}, \citenamefont {Massar}, \citenamefont
  {de~la Giroday}, \citenamefont {Matsukevich}, \citenamefont {Maunz},
  \citenamefont {Olmschenk}, \citenamefont {Hayes}, \citenamefont {Luo},
  \citenamefont {Manning},\ and\ \citenamefont {Monroe}}]{pironio2010}%
  \BibitemOpen
  \bibfield  {author} {\bibinfo {author} {\bibfnamefont {S.}~\bibnamefont
  {Pironio}}, \bibinfo {author} {\bibfnamefont {A.}~\bibnamefont {Ac\'in}},
  \bibinfo {author} {\bibfnamefont {S.}~\bibnamefont {Massar}}, \bibinfo
  {author} {\bibfnamefont {A.~Boyer}\ \bibnamefont {de~la Giroday}}, \bibinfo
  {author} {\bibfnamefont {D.~N.}\ \bibnamefont {Matsukevich}}, \bibinfo
  {author} {\bibfnamefont {P.}~\bibnamefont {Maunz}}, \bibinfo {author}
  {\bibfnamefont {S.}~\bibnamefont {Olmschenk}}, \bibinfo {author}
  {\bibfnamefont {D.}~\bibnamefont {Hayes}}, \bibinfo {author} {\bibfnamefont
  {L.}~\bibnamefont {Luo}}, \bibinfo {author} {\bibfnamefont {T.~A.}\
  \bibnamefont {Manning}}, \ and\ \bibinfo {author} {\bibfnamefont
  {C.}~\bibnamefont {Monroe}},\ }\bibfield  {title} {\enquote {\bibinfo {title}
  {Random numbers certified by bell's theorem},}\ }\href {\doibase
  doi:10.1038/nature09008} {\bibfield  {journal} {\bibinfo  {journal} {Nature}\
  }\textbf {\bibinfo {volume} {464}},\ \bibinfo {pages} {1021--1024} (\bibinfo
  {year} {2010})}\BibitemShut {NoStop}%
\bibitem [{\citenamefont {Acin}\ and\ \citenamefont
  {Masanes}(2016)}]{Acin2016}%
  \BibitemOpen
  \bibfield  {author} {\bibinfo {author} {\bibfnamefont {A.}~\bibnamefont
  {Acin}}\ and\ \bibinfo {author} {\bibfnamefont {Ll.}\ \bibnamefont
  {Masanes}},\ }\bibfield  {title} {\enquote {\bibinfo {title} {Certified
  randomness in quantum physics},}\ }\href {\doibase doi:10.1038/nature20119}
  {\bibfield  {journal} {\bibinfo  {journal} {Nature}\ }\textbf {\bibinfo
  {volume} {540}},\ \bibinfo {pages} {213} (\bibinfo {year}
  {2016})}\BibitemShut {NoStop}%
\bibitem [{\citenamefont {Christensen}\ \emph {et~al.}(2013)\citenamefont
  {Christensen}, \citenamefont {McCusker}, \citenamefont {Altepeter},
  \citenamefont {Calkins}, \citenamefont {Gerrits}, \citenamefont {Lita},
  \citenamefont {Miller}, \citenamefont {Shalm}, \citenamefont {Zhang},
  \citenamefont {Nam}, \citenamefont {Brunner}, \citenamefont {Lim},
  \citenamefont {Gisin},\ and\ \citenamefont {Kwiat}}]{christensen2013}%
  \BibitemOpen
  \bibfield  {author} {\bibinfo {author} {\bibfnamefont {B.~G.}\ \bibnamefont
  {Christensen}}, \bibinfo {author} {\bibfnamefont {K.~T.}\ \bibnamefont
  {McCusker}}, \bibinfo {author} {\bibfnamefont {J.~B.}\ \bibnamefont
  {Altepeter}}, \bibinfo {author} {\bibfnamefont {B.}~\bibnamefont {Calkins}},
  \bibinfo {author} {\bibfnamefont {T.}~\bibnamefont {Gerrits}}, \bibinfo
  {author} {\bibfnamefont {A.~E.}\ \bibnamefont {Lita}}, \bibinfo {author}
  {\bibfnamefont {A.}~\bibnamefont {Miller}}, \bibinfo {author} {\bibfnamefont
  {L.~K.}\ \bibnamefont {Shalm}}, \bibinfo {author} {\bibfnamefont
  {Y.}~\bibnamefont {Zhang}}, \bibinfo {author} {\bibfnamefont {S.~W.}\
  \bibnamefont {Nam}}, \bibinfo {author} {\bibfnamefont {N.}~\bibnamefont
  {Brunner}}, \bibinfo {author} {\bibfnamefont {C.~C.~W.}\ \bibnamefont {Lim}},
  \bibinfo {author} {\bibfnamefont {N.}~\bibnamefont {Gisin}}, \ and\ \bibinfo
  {author} {\bibfnamefont {P.~G.}\ \bibnamefont {Kwiat}},\ }\bibfield  {title}
  {\enquote {\bibinfo {title} {Detection-loophole-free test of quantum
  nonlocality, and applications},}\ }\href {\doibase
  10.1103/PhysRevLett.111.130406} {\bibfield  {journal} {\bibinfo  {journal}
  {Phys. Rev. Lett.}\ }\textbf {\bibinfo {volume} {111}},\ \bibinfo {pages}
  {130406} (\bibinfo {year} {2013})}\BibitemShut {NoStop}%
\bibitem [{\citenamefont {Paw\l{}owski}\ and\ \citenamefont
  {Brunner}(2011)}]{pawlowski2011}%
  \BibitemOpen
  \bibfield  {author} {\bibinfo {author} {\bibfnamefont {M.}~\bibnamefont
  {Paw\l{}owski}}\ and\ \bibinfo {author} {\bibfnamefont {N.}~\bibnamefont
  {Brunner}},\ }\bibfield  {title} {\enquote {\bibinfo {title}
  {Semi-device-independent security of one-way quantum key distribution},}\
  }\href {\doibase 10.1103/PhysRevA.84.010302} {\bibfield  {journal} {\bibinfo
  {journal} {Phys. Rev. A}\ }\textbf {\bibinfo {volume} {84}},\ \bibinfo
  {pages} {010302} (\bibinfo {year} {2011})}\BibitemShut {NoStop}%
\bibitem [{\citenamefont {Li}\ \emph {et~al.}(2011)\citenamefont {Li},
  \citenamefont {Yin}, \citenamefont {Wu}, \citenamefont {Zou}, \citenamefont
  {Wang}, \citenamefont {Chen}, \citenamefont {Guo},\ and\ \citenamefont
  {Han}}]{li2011}%
  \BibitemOpen
  \bibfield  {author} {\bibinfo {author} {\bibfnamefont {Hong-Wei}\
  \bibnamefont {Li}}, \bibinfo {author} {\bibfnamefont {Zhen-Qiang}\
  \bibnamefont {Yin}}, \bibinfo {author} {\bibfnamefont {Yu-Chun}\ \bibnamefont
  {Wu}}, \bibinfo {author} {\bibfnamefont {Xu-Bo}\ \bibnamefont {Zou}},
  \bibinfo {author} {\bibfnamefont {Shuang}\ \bibnamefont {Wang}}, \bibinfo
  {author} {\bibfnamefont {Wei}\ \bibnamefont {Chen}}, \bibinfo {author}
  {\bibfnamefont {Guang-Can}\ \bibnamefont {Guo}}, \ and\ \bibinfo {author}
  {\bibfnamefont {Zheng-Fu}\ \bibnamefont {Han}},\ }\bibfield  {title}
  {\enquote {\bibinfo {title} {Semi-device-independent random-number expansion
  without entanglement},}\ }\href {\doibase 10.1103/PhysRevA.84.034301}
  {\bibfield  {journal} {\bibinfo  {journal} {Phys. Rev. A}\ }\textbf {\bibinfo
  {volume} {84}},\ \bibinfo {pages} {034301} (\bibinfo {year}
  {2011})}\BibitemShut {NoStop}%
\bibitem [{\citenamefont {Li}\ \emph {et~al.}(2012)\citenamefont {Li},
  \citenamefont {Paw\l{}owski}, \citenamefont {Yin}, \citenamefont {Guo},\ and\
  \citenamefont {Han}}]{li2012}%
  \BibitemOpen
  \bibfield  {author} {\bibinfo {author} {\bibfnamefont {Hong-Wei}\
  \bibnamefont {Li}}, \bibinfo {author} {\bibfnamefont {Marcin}\ \bibnamefont
  {Paw\l{}owski}}, \bibinfo {author} {\bibfnamefont {Zhen-Qiang}\ \bibnamefont
  {Yin}}, \bibinfo {author} {\bibfnamefont {Guang-Can}\ \bibnamefont {Guo}}, \
  and\ \bibinfo {author} {\bibfnamefont {Zheng-Fu}\ \bibnamefont {Han}},\
  }\bibfield  {title} {\enquote {\bibinfo {title} {Semi-device-independent
  randomness certification using $n\rightarrow 1$ quantum random access
  codes},}\ }\href {\doibase 10.1103/PhysRevA.85.052308} {\bibfield  {journal}
  {\bibinfo  {journal} {Phys. Rev. A}\ }\textbf {\bibinfo {volume} {85}},\
  \bibinfo {pages} {052308} (\bibinfo {year} {2012})}\BibitemShut {NoStop}%
\bibitem [{\citenamefont {Bowles}\ \emph {et~al.}(2014)\citenamefont {Bowles},
  \citenamefont {Quintino},\ and\ \citenamefont {Brunner}}]{bowles2013}%
  \BibitemOpen
  \bibfield  {author} {\bibinfo {author} {\bibfnamefont {J.}~\bibnamefont
  {Bowles}}, \bibinfo {author} {\bibfnamefont {M.~T.}\ \bibnamefont
  {Quintino}}, \ and\ \bibinfo {author} {\bibfnamefont {N.}~\bibnamefont
  {Brunner}},\ }\bibfield  {title} {\enquote {\bibinfo {title} {Certifying the
  dimension of classical and quantum systems in a prepare-and-measure scenario
  with independent devices},}\ }\href {\doibase 10.1103/PhysRevLett.112.140407}
  {\bibfield  {journal} {\bibinfo  {journal} {Phys. Rev. Lett.}\ }\textbf
  {\bibinfo {volume} {112}},\ \bibinfo {pages} {140407} (\bibinfo {year}
  {2014})}\BibitemShut {NoStop}%
\bibitem [{\citenamefont {Woodhead}\ and\ \citenamefont
  {Pironio}(2015)}]{Woodhead2015}%
  \BibitemOpen
  \bibfield  {author} {\bibinfo {author} {\bibfnamefont {E.}~\bibnamefont
  {Woodhead}}\ and\ \bibinfo {author} {\bibfnamefont {S.}~\bibnamefont
  {Pironio}},\ }\bibfield  {title} {\enquote {\bibinfo {title} {Secrecy in
  prepare-and-measure clauser-horne-shimony-holt tests with a qubit bound},}\
  }\href {\doibase 10.1103/PhysRevLett.115.150501} {\bibfield  {journal}
  {\bibinfo  {journal} {Phys. Rev. Lett.}\ }\textbf {\bibinfo {volume} {115}},\
  \bibinfo {pages} {150501} (\bibinfo {year} {2015})}\BibitemShut {NoStop}%
\bibitem [{\citenamefont {Lunghi}\ \emph {et~al.}(2015)\citenamefont {Lunghi},
  \citenamefont {Brask}, \citenamefont {Lim}, \citenamefont {Lavigne},
  \citenamefont {Bowles}, \citenamefont {Martin}, \citenamefont {Zbinden},\
  and\ \citenamefont {Brunner}}]{lunghi2015}%
  \BibitemOpen
  \bibfield  {author} {\bibinfo {author} {\bibfnamefont {T.}~\bibnamefont
  {Lunghi}}, \bibinfo {author} {\bibfnamefont {J.~B.}\ \bibnamefont {Brask}},
  \bibinfo {author} {\bibfnamefont {C.~C.~W.}\ \bibnamefont {Lim}}, \bibinfo
  {author} {\bibfnamefont {Q.}~\bibnamefont {Lavigne}}, \bibinfo {author}
  {\bibfnamefont {J.}~\bibnamefont {Bowles}}, \bibinfo {author} {\bibfnamefont
  {A.}~\bibnamefont {Martin}}, \bibinfo {author} {\bibfnamefont
  {H.}~\bibnamefont {Zbinden}}, \ and\ \bibinfo {author} {\bibfnamefont
  {N.}~\bibnamefont {Brunner}},\ }\bibfield  {title} {\enquote {\bibinfo
  {title} {Self-testing quantum random number generator},}\ }\href {\doibase
  10.1103/PhysRevLett.114.150501} {\bibfield  {journal} {\bibinfo  {journal}
  {Phys. Rev. Lett.}\ }\textbf {\bibinfo {volume} {114}},\ \bibinfo {pages}
  {150501} (\bibinfo {year} {2015})}\BibitemShut {NoStop}%
\bibitem [{\citenamefont {Ca{\~n}as}\ \emph {et~al.}(2014)\citenamefont
  {Ca{\~n}as}, \citenamefont {Cari{\~n}e}, \citenamefont {G{\'o}mez},
  \citenamefont {Barra}, \citenamefont {Cabello}, \citenamefont {Xavier},
  \citenamefont {Lima},\ and\ \citenamefont {Paw{\l}owski}}]{canas2014}%
  \BibitemOpen
  \bibfield  {author} {\bibinfo {author} {\bibfnamefont {G.}~\bibnamefont
  {Ca{\~n}as}}, \bibinfo {author} {\bibfnamefont {J.}~\bibnamefont
  {Cari{\~n}e}}, \bibinfo {author} {\bibfnamefont {E.~S.}\ \bibnamefont
  {G{\'o}mez}}, \bibinfo {author} {\bibfnamefont {J.~F.}\ \bibnamefont
  {Barra}}, \bibinfo {author} {\bibfnamefont {A.}~\bibnamefont {Cabello}},
  \bibinfo {author} {\bibfnamefont {G.~B.}\ \bibnamefont {Xavier}}, \bibinfo
  {author} {\bibfnamefont {G.}~\bibnamefont {Lima}}, \ and\ \bibinfo {author}
  {\bibfnamefont {M.}~\bibnamefont {Paw{\l}owski}},\ }\bibfield  {title}
  {\enquote {\bibinfo {title} {Experimental quantum randomness generation
  invulnerable to the detection loophole},}\ }\href
  {https://arxiv.org/abs/1410.3443} {\bibfield  {journal} {\bibinfo  {journal}
  {arXiv:1410.3443}\ } (\bibinfo {year} {2014})}\BibitemShut {NoStop}%
\bibitem [{\citenamefont {Mironowicz}\ \emph {et~al.}(2016)\citenamefont
  {Mironowicz}, \citenamefont {Tavakoli}, \citenamefont {Hameedi},
  \citenamefont {Marques}, \citenamefont {Paw{\l}owski},\ and\ \citenamefont
  {Bourennane}}]{Mironowicz2016}%
  \BibitemOpen
  \bibfield  {author} {\bibinfo {author} {\bibfnamefont {P.}~\bibnamefont
  {Mironowicz}}, \bibinfo {author} {\bibfnamefont {A.}~\bibnamefont
  {Tavakoli}}, \bibinfo {author} {\bibfnamefont {A.}~\bibnamefont {Hameedi}},
  \bibinfo {author} {\bibfnamefont {B.}~\bibnamefont {Marques}}, \bibinfo
  {author} {\bibfnamefont {M.}~\bibnamefont {Paw{\l}owski}}, \ and\ \bibinfo
  {author} {\bibfnamefont {M.}~\bibnamefont {Bourennane}},\ }\bibfield  {title}
  {\enquote {\bibinfo {title} {Increased certification of semi-device
  independent random numbers using many inputs and more post-processing},}\
  }\href {http://stacks.iop.org/1367-2630/18/i=6/a=065004} {\bibfield
  {journal} {\bibinfo  {journal} {New J. Phys.}\ }\textbf {\bibinfo {volume}
  {18}},\ \bibinfo {pages} {065004} (\bibinfo {year} {2016})}\BibitemShut
  {NoStop}%
\bibitem [{\citenamefont {Vallone}\ \emph {et~al.}(2014)\citenamefont
  {Vallone}, \citenamefont {Marangon}, \citenamefont {Tomasin},\ and\
  \citenamefont {Villoresi}}]{vallone2014}%
  \BibitemOpen
  \bibfield  {author} {\bibinfo {author} {\bibfnamefont {G.}~\bibnamefont
  {Vallone}}, \bibinfo {author} {\bibfnamefont {D.~G.}\ \bibnamefont
  {Marangon}}, \bibinfo {author} {\bibfnamefont {M.}~\bibnamefont {Tomasin}}, \
  and\ \bibinfo {author} {\bibfnamefont {P.}~\bibnamefont {Villoresi}},\
  }\bibfield  {title} {\enquote {\bibinfo {title} {{Quantum randomness
  certified by the uncertainty principle}},}\ }\href {\doibase
  10.1103/PhysRevA.90.052327} {\bibfield  {journal} {\bibinfo  {journal} {Phys.
  Rev. A}\ }\textbf {\bibinfo {volume} {90}},\ \bibinfo {pages} {052327}
  (\bibinfo {year} {2014})}\BibitemShut {NoStop}%
\bibitem [{\citenamefont {Marangon}\ \emph {et~al.}(2017)\citenamefont
  {Marangon}, \citenamefont {Vallone},\ and\ \citenamefont
  {Villoresi}}]{Marangon2017}%
  \BibitemOpen
  \bibfield  {author} {\bibinfo {author} {\bibfnamefont {D.~G.}\ \bibnamefont
  {Marangon}}, \bibinfo {author} {\bibfnamefont {G.}~\bibnamefont {Vallone}}, \
  and\ \bibinfo {author} {\bibfnamefont {P.}~\bibnamefont {Villoresi}},\
  }\bibfield  {title} {\enquote {\bibinfo {title} {Source-device-independent
  ultrafast quantum random number generation},}\ }\href {\doibase
  10.1103/PhysRevLett.118.060503} {\bibfield  {journal} {\bibinfo  {journal}
  {Phys. Rev. Lett.}\ }\textbf {\bibinfo {volume} {118}},\ \bibinfo {pages}
  {060503} (\bibinfo {year} {2017})}\BibitemShut {NoStop}%
\bibitem [{\citenamefont {Cao}\ \emph {et~al.}(2016)\citenamefont {Cao},
  \citenamefont {Zhou}, \citenamefont {Yuan},\ and\ \citenamefont
  {Ma}}]{Cao2016}%
  \BibitemOpen
  \bibfield  {author} {\bibinfo {author} {\bibfnamefont {Z.}~\bibnamefont
  {Cao}}, \bibinfo {author} {\bibfnamefont {H.}~\bibnamefont {Zhou}}, \bibinfo
  {author} {\bibfnamefont {X.}~\bibnamefont {Yuan}}, \ and\ \bibinfo {author}
  {\bibfnamefont {X.}~\bibnamefont {Ma}},\ }\bibfield  {title} {\enquote
  {\bibinfo {title} {{Source-Independent Quantum Random Number Generation}},}\
  }\href {\doibase 10.1103/PhysRevX.6.011020} {\bibfield  {journal} {\bibinfo
  {journal} {Phys. Rev. X}\ }\textbf {\bibinfo {volume} {6}},\ \bibinfo {pages}
  {011020} (\bibinfo {year} {2016})}\BibitemShut {NoStop}%
\bibitem [{\citenamefont {Xu}\ \emph {et~al.}(2016)\citenamefont {Xu},
  \citenamefont {Shapiro},\ and\ \citenamefont {Wong}}]{Xu2016}%
  \BibitemOpen
  \bibfield  {author} {\bibinfo {author} {\bibfnamefont {Feihu}\ \bibnamefont
  {Xu}}, \bibinfo {author} {\bibfnamefont {Jeffrey~H.}\ \bibnamefont
  {Shapiro}}, \ and\ \bibinfo {author} {\bibfnamefont {Franco N.~C.}\
  \bibnamefont {Wong}},\ }\bibfield  {title} {\enquote {\bibinfo {title}
  {Experimental fast quantum random number generation using high-dimensional
  entanglement with entropy monitoring},}\ }\href {\doibase
  10.1364/OPTICA.3.001266} {\bibfield  {journal} {\bibinfo  {journal} {Optica}\
  }\textbf {\bibinfo {volume} {3}},\ \bibinfo {pages} {1266--1269} (\bibinfo
  {year} {2016})}\BibitemShut {NoStop}%
\bibitem [{\citenamefont {Cao}\ \emph {et~al.}(2015)\citenamefont {Cao},
  \citenamefont {Zhou},\ and\ \citenamefont {Ma}}]{Cao2015}%
  \BibitemOpen
  \bibfield  {author} {\bibinfo {author} {\bibfnamefont {Z.}~\bibnamefont
  {Cao}}, \bibinfo {author} {\bibfnamefont {H.}~\bibnamefont {Zhou}}, \ and\
  \bibinfo {author} {\bibfnamefont {X.}~\bibnamefont {Ma}},\ }\bibfield
  {title} {\enquote {\bibinfo {title} {{Loss-tolerant
  measurement-device-independent quantum random number generation}},}\ }\href
  {\doibase 10.1088/1367-2630/17/12/125011} {\bibfield  {journal} {\bibinfo
  {journal} {New J. Phys.}\ }\textbf {\bibinfo {volume} {17}},\ \bibinfo
  {pages} {125011} (\bibinfo {year} {2015})}\BibitemShut {NoStop}%
\bibitem [{\citenamefont {Ivanovic}(1987)}]{Ivanovic1987}%
  \BibitemOpen
  \bibfield  {author} {\bibinfo {author} {\bibfnamefont {I.D.}\ \bibnamefont
  {Ivanovic}},\ }\bibfield  {title} {\enquote {\bibinfo {title} {How to
  differentiate between non-orthogonal states},}\ }\href {\doibase
  http://dx.doi.org/10.1016/0375-9601(87)90222-2} {\bibfield  {journal}
  {\bibinfo  {journal} {Phys. Lett. A}\ }\textbf {\bibinfo {volume} {123}},\
  \bibinfo {pages} {257 -- 259} (\bibinfo {year} {1987})}\BibitemShut {NoStop}%
\bibitem [{\citenamefont {Dieks}(1988)}]{Dieks1988}%
  \BibitemOpen
  \bibfield  {author} {\bibinfo {author} {\bibfnamefont {D.}~\bibnamefont
  {Dieks}},\ }\bibfield  {title} {\enquote {\bibinfo {title} {Overlap and
  distinguishability of quantum states},}\ }\href {\doibase
  http://dx.doi.org/10.1016/0375-9601(88)90840-7} {\bibfield  {journal}
  {\bibinfo  {journal} {Phys. Lett. A}\ }\textbf {\bibinfo {volume} {126}},\
  \bibinfo {pages} {303 -- 306} (\bibinfo {year} {1988})}\BibitemShut {NoStop}%
\bibitem [{\citenamefont {Peres}(1988)}]{Peres1988}%
  \BibitemOpen
  \bibfield  {author} {\bibinfo {author} {\bibfnamefont {A.}~\bibnamefont
  {Peres}},\ }\bibfield  {title} {\enquote {\bibinfo {title} {How to
  differentiate between non-orthogonal states},}\ }\href {\doibase
  http://dx.doi.org/10.1016/0375-9601(88)91034-1} {\bibfield  {journal}
  {\bibinfo  {journal} {Phys. Lett. A}\ }\textbf {\bibinfo {volume} {128}},\
  \bibinfo {pages} {19 --} (\bibinfo {year} {1988})}\BibitemShut {NoStop}%
\bibitem [{\citenamefont {Chefles}(2000)}]{Chefles2000}%
  \BibitemOpen
  \bibfield  {author} {\bibinfo {author} {\bibfnamefont {A.}~\bibnamefont
  {Chefles}},\ }\bibfield  {title} {\enquote {\bibinfo {title} {{Quantum state
  discrimination}},}\ }\href {\doibase 10.1080/00107510010002599} {\bibfield
  {journal} {\bibinfo  {journal} {Contemp. Phys.}\ }\textbf {\bibinfo {volume}
  {41}},\ \bibinfo {pages} {401--424} (\bibinfo {year} {2000})}\BibitemShut
  {NoStop}%
\bibitem [{\citenamefont {Barnett}\ and\ \citenamefont
  {Croke}(2009)}]{Barnett2009}%
  \BibitemOpen
  \bibfield  {author} {\bibinfo {author} {\bibfnamefont {S.~M.}\ \bibnamefont
  {Barnett}}\ and\ \bibinfo {author} {\bibfnamefont {S.}~\bibnamefont
  {Croke}},\ }\bibfield  {title} {\enquote {\bibinfo {title} {{Quantum state
  discrimination}},}\ }\href {\doibase 10.1364/AOP.1.000238} {\bibfield
  {journal} {\bibinfo  {journal} {Adv. Opt. Photonics}\ }\textbf {\bibinfo
  {volume} {1}},\ \bibinfo {pages} {238} (\bibinfo {year} {2009})}\BibitemShut
  {NoStop}%
\bibitem [{Note1()}]{Note1}%
  \BibitemOpen
  \bibinfo {note} {Note that this bound assumes $\delta \geq 1/2$. If $\delta <
  1/2$, then the outcome will be most likely be conclusive, and therefore we
  have that $p_g \leqslant 1- \delta $.}\BibitemShut {Stop}%
\bibitem [{\citenamefont {Konig}\ \emph {et~al.}(2009)\citenamefont {Konig},
  \citenamefont {Renner},\ and\ \citenamefont {Schaffner}}]{Konig2009}%
  \BibitemOpen
  \bibfield  {author} {\bibinfo {author} {\bibfnamefont {R.}~\bibnamefont
  {Konig}}, \bibinfo {author} {\bibfnamefont {R.}~\bibnamefont {Renner}}, \
  and\ \bibinfo {author} {\bibfnamefont {C.}~\bibnamefont {Schaffner}},\
  }\bibfield  {title} {\enquote {\bibinfo {title} {{The Operational Meaning of
  Min- and Max-Entropy}},}\ }\href {\doibase 10.1109/TIT.2009.2025545}
  {\bibfield  {journal} {\bibinfo  {journal} {IEEE Trans. Inf. Theory}\
  }\textbf {\bibinfo {volume} {55}},\ \bibinfo {pages} {4337--4347} (\bibinfo
  {year} {2009})}\BibitemShut {NoStop}%
\bibitem [{Note2()}]{Note2}%
  \BibitemOpen
  \bibinfo {note} {Note though, that the assumption cannot be relaxed to a
  bound on the average overlap over many rounds, as there is then a classical
  strategy reproducing the USD probability distribution.}\BibitemShut {Stop}%
\bibitem [{\citenamefont {Stucki}\ \emph {et~al.}(2005)\citenamefont {Stucki},
  \citenamefont {Brunner}, \citenamefont {Gisin}, \citenamefont {Scarani},\
  and\ \citenamefont {Zbinden}}]{stucki2005}%
  \BibitemOpen
  \bibfield  {author} {\bibinfo {author} {\bibfnamefont {D.}~\bibnamefont
  {Stucki}}, \bibinfo {author} {\bibfnamefont {N.}~\bibnamefont {Brunner}},
  \bibinfo {author} {\bibfnamefont {N.}~\bibnamefont {Gisin}}, \bibinfo
  {author} {\bibfnamefont {V.}~\bibnamefont {Scarani}}, \ and\ \bibinfo
  {author} {\bibfnamefont {H.}~\bibnamefont {Zbinden}},\ }\bibfield  {title}
  {\enquote {\bibinfo {title} {Fast and simple one-way quantum key
  distribution},}\ }\href {\doibase 10.1063/1.2126792} {\bibfield  {journal}
  {\bibinfo  {journal} {Appl. Phys. Lett.}\ }\textbf {\bibinfo {volume} {87}},\
  \bibinfo {eid} {194108} (\bibinfo {year} {2005})}\BibitemShut {NoStop}%
\bibitem [{\citenamefont {Huttner}\ \emph {et~al.}(1996)\citenamefont
  {Huttner}, \citenamefont {Muller}, \citenamefont {Gautier}, \citenamefont
  {Zbinden},\ and\ \citenamefont {Gisin}}]{Huttner1996}%
  \BibitemOpen
  \bibfield  {author} {\bibinfo {author} {\bibfnamefont {B.}~\bibnamefont
  {Huttner}}, \bibinfo {author} {\bibfnamefont {A.}~\bibnamefont {Muller}},
  \bibinfo {author} {\bibfnamefont {J.~D.}\ \bibnamefont {Gautier}}, \bibinfo
  {author} {\bibfnamefont {H.}~\bibnamefont {Zbinden}}, \ and\ \bibinfo
  {author} {\bibfnamefont {N.}~\bibnamefont {Gisin}},\ }\bibfield  {title}
  {\enquote {\bibinfo {title} {Unambiguous quantum measurement of nonorthogonal
  states},}\ }\href {\doibase 10.1103/PhysRevA.54.3783} {\bibfield  {journal}
  {\bibinfo  {journal} {Phys. Rev. A}\ }\textbf {\bibinfo {volume} {54}},\
  \bibinfo {pages} {3783--3789} (\bibinfo {year} {1996})}\BibitemShut {NoStop}%
\bibitem [{\citenamefont {Hoeffding}(1963)}]{hoeffding1963}%
  \BibitemOpen
  \bibfield  {author} {\bibinfo {author} {\bibfnamefont {W.}~\bibnamefont
  {Hoeffding}},\ }\bibfield  {title} {\enquote {\bibinfo {title} {Probability
  inequalities for sums of bounded random variables},}\ }\href {\doibase
  10.1080/01621459.1963.10500830} {\bibfield  {journal} {\bibinfo  {journal}
  {J. Am. Stat. Assoc.}\ }\textbf {\bibinfo {volume} {58}},\ \bibinfo {pages}
  {13--30} (\bibinfo {year} {1963})}\BibitemShut {NoStop}%
\bibitem [{Note3()}]{Note3}%
  \BibitemOpen
  \bibinfo {note} {The two peaks of the top curve on the left plot in Fig.~\ref
  {fig_hmin} arise from two different measurement strategies maximising the
  guessing probabilities, while reproducing the observed data, in different
  loss regimes. Disregarding finite size effects, for low loss, a USD strategy
  is optimal, and fixing the conclusive rate at 1/2 requires a choice of
  $|\alpha |^2 = \protect \qopname \relax o{log}(2)/\eta $. For increasing
  loss, a mixture of strategies where only one state is identified (i.e. one of
  the outputs 0 or 1 has vanishing probability) becomes optimal. In this case
  the inconclusive rate is $\delta ^2$ and the optimal $|\alpha |^2 = \protect
  \qopname \relax o{log}(\protect \sqrt {2})$.}\BibitemShut {Stop}%
\bibitem [{\citenamefont {Todoroki}\ and\ \citenamefont
  {Inoue}(2004)}]{Todoroki2004}%
  \BibitemOpen
  \bibfield  {author} {\bibinfo {author} {\bibfnamefont {Shin-ichi}\
  \bibnamefont {Todoroki}}\ and\ \bibinfo {author} {\bibfnamefont {Satoru}\
  \bibnamefont {Inoue}},\ }\bibfield  {title} {\enquote {\bibinfo {title}
  {{Optical Fuse by Carbon-Coated TeO 2 Glass Segment Inserted in Silica Glass
  Optical Fiber Circuit}},}\ }\href {\doibase 10.1143/JJAP.43.L256} {\bibfield
  {journal} {\bibinfo  {journal} {Jpn. J. Appl. Phys.}\ }\textbf {\bibinfo
  {volume} {43}},\ \bibinfo {pages} {L256--L257} (\bibinfo {year}
  {2004})}\BibitemShut {NoStop}%
\bibitem [{QRN()}]{QRNG}%
  \BibitemOpen
  \href@noop {} {\enquote {\bibinfo {title} {http://www.idquantique.com \&
  http://www.picoquant.com},}\ }\BibitemShut {NoStop}%
\bibitem [{\citenamefont {Himbeeck}\ \emph {et~al.}(2016)\citenamefont
  {Himbeeck}, \citenamefont {Woodhead}, \citenamefont {Garcia-Patron},
  \citenamefont {Cerf},\ and\ \citenamefont {Pironio}}]{Himbeck}%
  \BibitemOpen
  \bibfield  {author} {\bibinfo {author} {\bibfnamefont {T.~Van}\ \bibnamefont
  {Himbeeck}}, \bibinfo {author} {\bibfnamefont {E.}~\bibnamefont {Woodhead}},
  \bibinfo {author} {\bibfnamefont {R.S.}\ \bibnamefont {Garcia-Patron}},
  \bibinfo {author} {\bibfnamefont {N.}~\bibnamefont {Cerf}}, \ and\ \bibinfo
  {author} {\bibfnamefont {S.}~\bibnamefont {Pironio}},\ }\bibfield  {title}
  {\enquote {\bibinfo {title} {Semi-device-independent framework based on
  natural physical assumptions},}\ }\href {https://arxiv.org/abs/1612.06828}
  {\bibfield  {journal} {\bibinfo  {journal} {arXiv [quant-ph]}\ ,\ \bibinfo
  {pages} {1612.06828}} (\bibinfo {year} {2016})}\BibitemShut {NoStop}%
\bibitem [{\citenamefont {Bancal}\ \emph {et~al.}(2014)\citenamefont {Bancal},
  \citenamefont {Sheridan},\ and\ \citenamefont {Scarani}}]{bancal2014}%
  \BibitemOpen
  \bibfield  {author} {\bibinfo {author} {\bibfnamefont {Jean-Daniel}\
  \bibnamefont {Bancal}}, \bibinfo {author} {\bibfnamefont {Lana}\ \bibnamefont
  {Sheridan}}, \ and\ \bibinfo {author} {\bibfnamefont {Valerio}\ \bibnamefont
  {Scarani}},\ }\bibfield  {title} {\enquote {\bibinfo {title} {More randomness
  from the same data},}\ }\href {https://doi.org/10.1088/1367-2630/16/3/033011}
  {\bibfield  {journal} {\bibinfo  {journal} {New J. Phys.}\ }\textbf {\bibinfo
  {volume} {16}},\ \bibinfo {pages} {033011} (\bibinfo {year}
  {2014})}\BibitemShut {NoStop}%
\end{thebibliography}%

\appendix

\section{Bounding $p_g$ by semidefinite programing}
\label{sec.sdp}

In this Appendix, we show how the guessing probability can be bounded via SDP. We discuss both the primal and dual programs. We start our analysis by assuming a fixed overlap $|\braket{\psi_0}{\psi_1}| = \delta$ between the two prepared states, and show in the end that this is general, i.e. that the case $|\braket{\psi_0}{\psi_1}| = \Delta > \delta$ is covered.

\subsection{Primal}

For a fixed overlap $|\braket{\psi_0}{\psi_1}| = \delta$ and given data $p(b|x)$, the guessing probability is bounded by the maximisation over all measurement strategies and their distribution, reproducing the data. Assuming that the inputs are balanced, $p(x) = 1/2$, and denoting the distribution of measurement strategies by $q_\lambda = p(\lambda)$ and the density matrices $\rho_x = \ket{\psi_x} 
\bra{\psi_x}$, we have that
\begin{equation}
p_g \leqslant \frac{1}{2} \max_{q_\lambda, \Pi^\lambda_b} \sum^1_{x=0} \sum_\lambda q_\lambda \max\{\Tr\left[\rho_x \Pi^\lambda_{\o} \right], 1 -\Tr\left[\rho_x \Pi^\lambda_{\o} \right]\},
\end{equation}
with the constraint that the data is reproduced, i.e. that $\sum_\lambda q_\lambda \Tr\left[\rho_x \Pi^\lambda_{\o} \right]  = p(b|x)$. We note that, although it looks like the above expression depends on the states $\rho_x$, this is not actually the case, as the trace is invariant under unitary transformations. Furthermore, since there are just two states we can restrict to a 2-dimensional Hilbert space without loss of generality. Hence, we can take the two states to be $\ket{\psi_0} = \ket 0$ and $\ket{\psi_1} = \delta \ket 0 + \sqrt{1-\delta} \ket 1$ in some basis $\{\ket{0}, \ket{1}\} $. It is then clear that the maximum depends only on $\delta$ and the observed data $ p(b|x)$.

A priori, the number of measurement strategies is unbounded. However, following \cite{bancal2014}, all strategies for which the inner maximization occurs for the same term can be grouped together. It is then sufficient to consider four different measurement strategies corresponding to the max occurring for the first or second term for each $x$, and one can remove the inner maximization without loss of generality. We label these strategies by $(\lambda_0, \lambda_1)$ where $\lambda_x$ determines which term is maximal for the input $x$. We thus have four POVMs with elements $\Pi^{\lambda_0,\lambda_1}_b$.  Defining $\widetilde{\Pi}^{\lambda_0,\lambda_1}_c = \delta_{c,0} \Pi^{\lambda_0,\lambda_1}_{\o} +  \delta_{c,1} \left( \mathbbm{1} - \Pi^{\lambda_0,\lambda_1}_{\o} \right)$, the bound can be written
\begin{equation}
p_g \leqslant \frac{1}{2} \max_{q_{\lambda_0, \lambda_1}, \Pi^{\lambda_0, \lambda_1}_b} \sum^1_{x=0} \sum^1_{\lambda_0, \lambda_1 = 0} q_{\lambda_0,\lambda_1} \Tr\left[\rho_x \widetilde{\Pi}^{\lambda_0,\lambda_1}_{\lambda_x} \right].
\end{equation}
Finally, we absorb the weights $q_{\lambda_0, \lambda_1}$ into the POVM elements and define $M_b^{\lambda_0, \lambda_1} = q_{\lambda_0, \lambda_1} \Pi^{\lambda_0, \lambda_1}_b $, and $\widetilde M_c^{\lambda_0, \lambda_1} = \delta_{c,0}M_{\o}^{\lambda_0, \lambda_1} + \delta_{c,1} \left( \mathbbm{1} - M_{\o}^{\lambda_0, \lambda_1} \right)$. With this, we arrive at a bound $p_g \leqslant \bar{p}_g$ which can be computed by semidefinite programming
\begin{equation}
\label{eq.pgprimal}
\bar{p}_g = \frac{1}{2} \max_{M^{\lambda_0, \lambda_1}_b} \sum^1_{x=0} \sum^1_{\lambda_0, \lambda_1 = 0} \Tr\left[ \rho_x  \widetilde M_{\lambda_x}^{\lambda_0, \lambda_1} \right],
\end{equation}
subject to the constraints that the $M_b^{\lambda_0, \lambda_1}$ be hermitian, positive semidefinite, sum to the identity, that they form a valid, subnormalised measurement for each $(\lambda_0, \lambda_1)$, and that the data is reproduced. That is
\begin{align}
M_b^{\lambda_0, \lambda_1} & = \left(M_b^{\lambda_0, \lambda_1}\right)^\dagger,\\
\label{eq.primalcons_positivity}
M_b^{\lambda_0, \lambda_1} & \geqslant 0, \\
\label{eq.primalcons_subnorm}
\sum_b M_b^{\lambda_0, \lambda_1} & = \frac{1}{2} \Tr \left[\sum_b M_b^{\lambda_0, \lambda_1}\right] \mathbbm{1},\\
\label{eq.primalcons_data}
\sum_{\lambda_0, \lambda_1} \Tr  \left[\rho_ xM_b^{\lambda_0, \lambda_1} \right] & = p(b|x).
\end{align}
Note that normalisation of $\rho_x$ and $p(b|x)$ together with conditions \eqref{eq.primalcons_subnorm} and \eqref{eq.primalcons_data} imply that $\sum_{b,\lambda_0, \lambda_1} \Tr  \left[M_b^{\lambda_0, \lambda_1} \right] = 2$. Since \eqref{eq.pgprimal} is linear in the $M_b^{\lambda_0, \lambda_1}$, and the constraints are semidefinite, the maximisation defines an SDP and can be solved efficiently, providing optimal bounds on $p_g$ for every given state overlap and observed data.

\subsection{Dual}

While the primal SDP above gives optimal bounds on the guessing probability for given observed data and a fixed state overlap, it is not practical to incorporate directly into the QRNG for several reasons. The first is speed. Every time the distribution $p(b|x)$ is updated based on the raw data, the SDP must be evaluated to update the bound. This evaluation typically takes on the order of a second, potentially slowing down the bit rate significantly. Second, experimentally the state overlap is not known exactly, but a lower bound can be established with high certainty. Hence, one would like a bound which is valid for any larger overlap. Third, since $p(b|x)$ is estimated from finite raw data, finite-size effects must be accounted for in the bound. It is not obvious how to incorporate this into the primal SDP in an efficient manner. 

Fortunately, all of these concerns can be addressed by using the dual SDP. A solution of the dual provides an upper bound on the solution of the primal, and hence on $p_g$. When the data $p(b|x)$ changes, a new bound can be found by evaluating a simple, linear function of $p(b|x)$ with no need to run the full SDP as long as $\delta$ is fixed. Furthermore, because the function is linear, finite-size effects can be incorporated straightforwardly. The bound can be shown to hold for any overlap $\Delta \geqslant \delta$, as discussed at the end of this section.

We now derive the dual SDP in a manner which makes it clear that it upper bounds the primal. For each of the constraints in \eqref{eq.primalcons_positivity}-\eqref{eq.primalcons_data} we introduce Lagrangian multipliers, respectively hermitian 2x2 matrices $G^{\lambda_0,\lambda_1}_b$, $H^{\lambda_0,\lambda_1}$, and real scalars $\nu_{bx}$. We define a Lagrangian function of the primal SDP variables and these new variables, given by
\begin{align}
\label{eq.lagrangian}
& \mathcal{L} = \nonumber \\
& \frac{1}{2} \sum_{x=0}^1 \sum_{\lambda_0,\lambda_1=0}^1 \Tr[\rho_x \left( \delta_{\lambda_x,0} M^{\lambda_0,\lambda_1}_{\o} + \delta_{\lambda_x,1}(\mathbbm{1}-M^{\lambda_0,\lambda_1}_{\o})\right)] \nonumber \\
& + \sum_{b,\lambda_0,\lambda_1} \Tr[G^{\lambda_0,\lambda_1}_b M^{\lambda_0,\lambda_1}_b] \\
& + \sum_{\lambda_0,\lambda_1} \Tr[H^{\lambda_0,\lambda_1} \sum_b ( M^{\lambda_0,\lambda_1}_b - \frac{1}{2}\Tr[M^{\lambda_0,\lambda_1}_b] \mathbbm{1} )] \nonumber \\
& + \sum_{x,b} \nu_{bx} (\sum_{\lambda_0,\lambda_1} \Tr[\rho_x M^{\lambda_0,\lambda_1}_b ] - p(b|x)) . \nonumber
\end{align}
We further define $\mathcal{S}$ to be the supremum of the Lagrangian over the primal SDP variables. That is
\begin{equation}
\mathcal{S} = \underset{M^{\lambda_0,\lambda_1}_b}{\sup} \mathcal{L} .
\end{equation}
For any particular solution $M^{\lambda_0,\lambda_1}_b$ of the primal SDP \eqref{eq.pgprimal}-\eqref{eq.primalcons_data}, the two last terms in the Lagrangian \eqref{eq.lagrangian} vanish, because the solution fulfills the constraints \eqref{eq.primalcons_subnorm}-\eqref{eq.primalcons_data}. Similarly, because of \eqref{eq.primalcons_positivity}, the second term in the Lagrangian is positive if the $G^{\lambda_0,\lambda_1}_b$ are restricted to be positive. The first term of the Lagrangian is the target function of the primal \eqref{eq.pgprimal}. It follows that $\mathcal{S}$ is an upper bound on the value of the primal, $\mathcal{S} \geqslant \bar{p}_g$, when $G^{\lambda_0,\lambda_1}_b \geqslant 0$, and thus also an upper bound on the guessing probability $\mathcal{S} \geqslant p_g$.

To get good bounds, we should minimise $\mathcal{S}$ over the Lagrangian multipliers. To this end, we first rewrite $\mathcal{S}$ in a more convenient form. We collect all terms which multiply the primal variables.
\begin{equation}
\label{eq.suprewrite}
\mathcal{S} = \underset{M^{\lambda_0,\lambda_1}_b}{\sup} \,\, \sum_{b,\lambda_0,\lambda_1} \Tr [M^{\lambda_0,\lambda_1}_b K^{\lambda_0,\lambda_1}_b] - \sum_{b,x} \nu_{bx} p(b|x) ,
\end{equation}
where
\begin{align}
\label{eq.Kexpr}
K^{\lambda_0,\lambda_1}_b & = \sum_x \rho_x ( \frac{1}{2}\delta_{\lambda_x,0}\delta_{b,\o} + \frac{1}{2}\delta_{\lambda_x,1}(1-\delta_{b,\o}) + \nu_{bx} ) \nonumber \\
& + G^{\lambda_0,\lambda_1}_b + H^{\lambda_0,\lambda_1} - \frac{1}{2}\Tr[H^{\lambda_0,\lambda_1}]\mathbbm{1} .
\end{align}
Since here the $M^{\lambda_0,\lambda_1}_b$ are not restricted to being positive, we see that the supremum in \eqref{eq.suprewrite} will be infinite, unless $K^{\lambda_0,\lambda_1}_b$ vanishes. Hence, to get good bounds on $p_g$ we must impose that $K^{\lambda_0,\lambda_1}_b=0$.  Since the operators $G^{\lambda_0,\lambda_1}_b$ are positive semidefinite but not otherwise restricted, this is equivalent to dropping $G^{\lambda_0,\lambda_1}_b$ from \eqref{eq.Kexpr} and requiring that the remaining expression is negative semidefinite. Using this, we finally arrive at our dual SDP
\begin{equation}
\label{eq.dual}
p_g^* = \underset{H^{\lambda_0,\lambda_1},\nu_{bx}}{\min} - \sum_{bx} \nu_{bx} p(b|x)
\end{equation}
subject to
\begin{align}
\label{eq.dualcons_herm}
H^{\lambda_0,\lambda_1} & = (H^{\lambda_0,\lambda_1})^\dagger , \\
\label{eq.dualcons_cons}
\sum_x \rho_x & ( \frac{1}{2}\delta_{\lambda_x,0}\delta_{b,\o} + \frac{1}{2}\delta_{\lambda_x,1}(1-\delta_{b,\o}) + \nu_{bx} ) \nonumber \\
& + H^{\lambda_0,\lambda_1} - \frac{1}{2}\Tr[H^{\lambda_0,\lambda_1}]\mathbbm{1} \leqslant 0 .
\end{align}
From the above, it should be clear that $p_g \leqslant \bar{p}_g \leqslant p_g^*$. We also see that the data $p(b|x)$ does not appear in the dual constraints \eqref{eq.dualcons_herm}-\eqref{eq.dualcons_cons}. This means that given one feasible dual solution (a set of $H^{\lambda_0,\lambda_1}$ and $\nu_{bx}$ fulfilling the constraints), valid bounds on $p_g$ can be computed for any data $p(b|x)$ by evaluating the right-hand-side of \eqref{eq.dual}. This is a simple linear function and can be evaluated very fast in practice. Furthermore, this form allows us to treat finite-size effects easily, as explained in the main text; see Eq. \eqref{eq:pg_finitesize}.

\begin{figure}[b!]
  \centering
  \includegraphics[width=0.8\columnwidth]{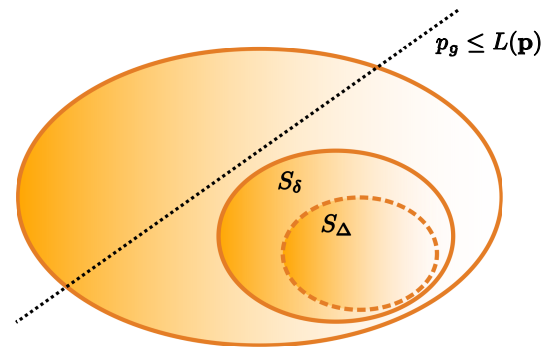}
  \caption{Larger set: space of all conditional distributions $p(b|x)$. Set $S_\delta$: distributions obtainable from states with overlap $\delta$. Set $S_\Delta$: distributions obtainable from states with overlap $\Delta > \delta$. A bound obtained from the dual SDP using overlap $\delta$ defines a hyperplane (dashed line) with $S_\delta$ on one side. To see that the bound holds for all $\Delta > \delta$, it is sufficient to realise that $S_\Delta \subseteq S_\delta$.}
  \label{fig.prob_space}
\end{figure}

The dual bound $p_g^*$ also remains valid when the overlap of the input states increases. To see this, consider the space of conditional distributions $p(b|x)$ thought of as vectors $\mathbf{p}$. A bound of the form
\begin{equation}
p_g \leqslant \sum_{bx} \nu_{bx} p(b|x) = L(\mathbf{p}),
\end{equation}
for fixed numbers $p_g$, $\nu_{bx}$ defines a hyperplane in this space, with all distributions $p(b|x)$ fulfilling the bound lying in one of the corresponding half spaces. Let us denote the set of all distributions which can be generated from a pair of pure states with overlap $\delta$ by $S_\delta$. It is easy to see that this set must be convex. We then have a picture as in \figref{fig.prob_space}. Since the bound on $p_g$ holds for all points in $S_\delta$, to show that it also holds for all $\Delta > \delta$, it is sufficient to show that $S_\Delta \subseteq S_\delta$, i.e.~that any distribution which can be obtained from two states with overlap $\Delta$ can also be obtained from two states with smaller overlap $\delta$.

This can be shown as follows. Consider two pure states $\ket{\psi_0}$, $\ket{\psi_1}$ with overlap $\Delta$. We add an ancilla system, and define states $\ket{\phi_x} = \ket{\psi_x}\ket{0}$, and $\ket{\varphi_0} = \ket{\psi_0}\ket{0}$, $\ket{\varphi_1} = \ket{\psi_1}\ket{s}$, where $\ket{0}$ is some fixed state and $\ket{s}$ a different state. Then $|\braket{\phi_0}{\phi_1}| = \Delta$, and $|\braket{\varphi_0}{\varphi_1}| = \Delta |\braket{0}{s}| = \delta$, where $\delta$ can be set to any value  $\leq \Delta$ by adjusting the overlap of the ancilla states $|\braket{0}{s}|$.

Now, any distribution $p(b|x) = \Tr[M_b\ket{\psi_x}\bra{\psi_x}]$ which can be obtained from the states $\ket{\psi_x}$ can clearly also be obtained from $\ket{\phi_x}$ by extending the POVM trivially, $p(b|x) = \Tr[(M_b\otimes\mathbbm{1})\ket{\phi_x}\bra{\phi_x}]$. However, the same POVM acting on the states $\ket{\varphi_x}$ will give the same distribution, because it is acting trivially on the ancilla, $p(b|x) = \Tr[(M_b\otimes\mathbbm{1})\ket{\varphi_x}\bra{\varphi_x}]$. Hence, for any distribution obtained from a POVM on a pair of pure states with overlap $\Delta$, there exists another pair of pure states with overlap $\delta$ and a POVM reproducing the distribution.

Finally, we observe that since we are working only with pairs of states, the ancilla is in fact unnecessary. Any $p(b|x)$ obtained from a pair of pure states can be obtained from a pair of qubit states (with the same overlap). Also, since any pair of pure qubit states is unitarily related to any other pair with the same overlap, it follows that \textit{any} pair with overlap $\delta$ can reproduce the measurement statistics from \textit{any} pair with overlap $\Delta \geq \delta$.

\end{document}